\documentclass[a4paper]{article}
\usepackage[a4paper,top=3cm,bottom=2cm,left=3cm,right=3cm,marginparwidth=1.75cm]{geometry}
\usepackage{amsfonts}
\usepackage{bm}
\usepackage{extarrows}
\usepackage{amssymb}
\usepackage{color}
\usepackage[all]{xy}
\usepackage{graphicx}
\usepackage{braket}
\usepackage{amsmath}
\usepackage{appendix}
\usepackage{graphicx}
\usepackage[colorinlistoftodos]{todonotes}
\usepackage{amsthm}
\usepackage{mathrsfs}
\usepackage{amssymb}
\usepackage{accents}
\usepackage{extarrows}
\usepackage{makecell}
\usepackage{authblk}
\usepackage{url}
\usepackage{hyperref}
\usepackage{cite}
\usepackage{eufrak}
\hypersetup{colorlinks,
            citecolor=black,
            linkcolor=black,
            urlcolor=green,
            pdftex}

\usepackage[english]{babel}
\usepackage[utf8x]{inputenc}
\usepackage[T1]{fontenc}

\newcommand{\vev}[1]{\left\langle #1 \right\rangle}

\usepackage[normalem]{ulem}

\title{Perelomov type coherent states of SO$(D+1)$ in all dimensional loop quantum gravity}
\author[1,3]{Gaoping Long \footnote{201731140005@mail.bnu.edu.cn}\thanks{corresponding author}}
\author[2]{Norbert Bodendorfer \footnote{norbert.bodendorfer@physik.uni-regensburg.de}}

\affil[1]{Department of Physics, South China University of Technology, Guangzhou 510641, China}
\affil[2]{Institute for Theoretical Physics, University of Regensburg, 93040 Regensburg, Germany}
\affil[3]{Department of Physics, Beijing Normal University, Beijing 100875, China}

\date{}

\begin{document}

\maketitle

\begin{abstract}
A comprehensive study of the application of SO$(D+1)$ coherent states of Perelomov type to loop quantum gravity in general spacetime dimensions $1+D\geq 3$ is given in this paper.
We focus on so-called simple representations of SO$(D+1)$ which solve the simplicity constraint acting on edges and the associated homogeneous harmonic function spaces.
With the harmonic function formulation, we study general properties of the coherent states such as the peakedness properties and the inner product. We also discuss the properties of geometric operators evaluated in the coherent states. In particular, we calculate the expectation value of the volume operator, and the results agree with the ones obtained from the classical label of the coherent states up to error terms which vanish in the limit of large representation labels $N$, i.e. the analogue of the large spin limit in standard $(1+3)$-dimensional loop quantum gravity.

\end{abstract}

\section{Introduction}

Loop quantum method serves a non-perturbative and background-independent way to explore the quantum theory of general relativity (GR), and it is used to construct the standard (1+3)-dimensional loop quantum gravity (LQG) successfully \cite{thiemann2007modern, rovelli2007quantum, Ashtekar2012Background, Han2005FUNDAMENTAL}. Meanwhile, superstring theory is also proposed as a quantum gravity theory in 10-dimensional spacetime. In addition to the quantum gravity part, superstring theory also includes extra quantum degrees of freedom which are expected to describe the quantum matter field. In fact, the extra-dimension idea is widely used to unify the gravity and the other three fundamental interactions in several theories, i.e., Kaluza-Klein theory. Inspired by these higher dimensional theories, it is therefore interesting to explore the loop quantum theory in arbitrary (1+D)-dimensional spacetime with $D>3$.


The first step to explore the all dimensional dimensional LQG is to reformulate arbitrary (1+D)-dimensional GR as connection dynamics theory. An alternative scheme was proposed
by Bodendorfer, Thiemann and Thurn \cite{ Bodendorfer:Ha, Bodendorfer:La, Bodendorfer:Qu, Bodendorfer:2011onthe, Bodendorfer:SgI}. In this scheme, the connection formulation is
achieved by extending the ADM phase space (1+D)-dimensional GR as a Yang-Mills phase space with
gauge group SO$(D+1)$, and extra Gaussian constraint and simplicity constraint which constitutes a first class constraint system with the vector constraint and scalar constraint are introduce to eliminate
the gauge degrees of freedom.
In addition to the usual constraints in standard (1+3)-dimensional LQG, the formulation of LQG in general spacetime dimensions includes the so-called simplicity constraints which enforce that the fluxes, which transform in the adjoint representation of SO$(D+1)$, are constructed from bi-vectors, i.e. $\pi^{aIJ} = 2 n^{[I} E^{a|J]}$, where $a,b = 1,\ldots,D$ are spatial tensor indices and $I,J = 1, \ldots, D+1$ are vector indices of SO$(D+1)$. $E^{aJ}$ is the analogue of the densitized triad and $n^I$ is an internal normal satisfying $n_I E^{aI} = 0$. The quantization of this SO$(D+1)$ gauge theory can be achieved by following the standard loop quantization methods, and the resulting all dimensional LQG is equipped with a kinematic Hilbert space given by the completion of the space of cylindrical functions on certain quantum configuration space, the basic operators acting on the cylindrical functions and corresponding quantum constraints. Especially, at the quantum level, the simplicity constraints are split into two distinct groups, the first acting on spin network edges and the second acting on vertices. The former are non-anomalous and easily solved by restricting the SO$(D+1)$-representations to so-called simple ones \cite{Bodendorfer:2011onthe}. The latter on the other hand are anomalous, a fact well known from earlier investigations in spin foam models, see e.g. \cite{FreidelBFDescriptionOf}\cite{RovelliBook2}. Imposing them strongly eliminates too many physical degrees of freedom and alternative strategies have to be developed, see e.g. \cite{Bodendorfer:2011onthe} for an approach using maximally commuting subsets.

Another choice to deal with this problem is to try to solve the anomalous constraints weakly, see e.g. \cite{Engle:2007LQGv, Engle:2007Fsv, Dupuis:2010RSC} for previous work and \cite{long2019coherent} for an application to SO$(D+1)$ Perelomov coherent states \cite{GeneralizedCoherentStates}. In order to achieve this, the properties of flux operators sandwiched between coherent states are needed. In previous work \cite{long2019coherent}, the basic peakedness property of such coherent state was discussed. It turned out to play the key role to weakly solve the quantum vertex simplicity constraints and to minimize the occuring errors.
It also turned out that the simple coherent intertwiner space \cite{Livine:2007Nsfv, Dupuis:2010RSC}, similar to work in $(1+3)$ dimensions \cite{Bianchi:2010Polyhedra}, can be regarded as the quantum space of the shape space of $D$-polytopes \cite{Bodendorfer:2013sja, long2019coherent}.

More generally, in LQG, the intrinsic spatial geometry is completely determined by the flux operators, so that simple coherent intertwiners are suitable candidates for coherent states in which a large class of intrinsic geometric operators may be sharply peaked. Based on this idea, the expectation values of the geometric operators in the states labelled by simple coherent intertwiners are expected to have minimal, or close to minimal, quantum uncertainties. However, the calculation of expectation values of geometric operators is usually much more complicated than the calculation for flux operators. On the one hand, this is due to the geometric operators not being simple polynomials in the fluxes. On the other hand, the group averaging introduced in the construction of the gauge invariant simple coherent intertwiners complicates matters. Hence, a more comprehensive study of the Perelomov coherent state of SO$(D+1)$ and simple coherent intertwiners is necessary.

For readers familiar with previous work in $(1+3)$ dimensions, let us mention that the SO$(D+1)$ coherent states of Perelomov type in the simple representation spaces satisfying the edge simplicity constraints are the higher dimensional extension of the SU$(2)$ coherent states of Perelomov type \cite{GeneralizedCoherentStates}, which are the coherent states for angular momentum in three-dimensional space. Similar to the SU$(2)$ case, the SO$(D+1)$ coherent states of Perelomov type are given by rotating the state $|N\mathbf{e}_1\rangle$ with an arbitrary element $g\in SO(D+1)$, where $|N\mathbf{e}_1\rangle$ is the state which corresponds to the highest weight vector $N\mathbf{e}_1$ in a simple representation space labelled by a non-negative integer $N$ \cite{Simon, vilenkin2013representation}. In addition to $N$, the final coherent states $|N,V\rangle$ are determined by a bi-vector $V$ which labels the equivalence class of the group elements that rotate $|N\mathbf{e}_1\rangle$ to $|N,V\rangle$. The SO$(D+1)$ coherent states of Perelomov type are expected to have a series of properties such as minimizing the Heisenberg uncertainty relation applied to flux operators. Besides, some other properties of SU$(2)$ coherent states are expected to be extendable to the SO$(D+1)$ case, such as the non-orthogonal property and the form of the inner product of two coherent states. This will be the topic of the first part of this paper.

This paper is organized as follows. In section 2, we will review the angular momentum theory in higher dimensions, which gives a more familiar realization of the quantum algebra of flux operators. Also, we will review the representation theory of SO$(D+1)$ in the harmonic function space and give a comprehensive study of the properties of the SO$(D+1)$ coherent states of Perelomov type in section 3. In section 4, we will discuss some corresponding properties of the spin network states which are labelled with simple coherent intertwiners in all dimensional LQG, as well as introduce some applications of these properties in the calculation of expectation value of geometric operators. In the final section 5, the conclusion of our results will be given. An appendix provides an error estimate and the details of discussion for some of our calculations.

\section{Quantum algebra of flux operators from a particle moving on a $D$-sphere}
For pedagogical purposes, we will review the phase space structure and quantum mechanics of a particle moving on the $D$-sphere as discussed in \cite{hall2002coherent} and compare it with the flux operators in LQG.
Consider the $D$-dimensional sphere $S^D$ with unit radius in $\mathbb{R}^{D+1}$ as the configuration space for a particle moving on it ($D\geq1$). The associated
phase space, the cotangent bundle $T^\ast(S^D)$ is given by
\begin{equation}
T^\ast(S^D)=\{(\bm{x},\bm{p})\in \mathbb{R}^{D+1}\times \mathbb{R}^{D+1}|\bm{x}\cdot\bm{x}=1, \bm{x}\cdot\bm{p}=0\},
\end{equation}
where $\bm{x}=(x_1,...,x_I,..,x_{D+1})$, $\bm{p}=(p_1,..,p_J,...p_{D+1})$ are vectors in $\mathbb{R}^{D+1}$, representing respectively the position and momentum of the particle. We can now define the angular momentum of the particle as $J_{IJ}:=x_Ip_J-x_Jp_I$, or alternatively, describe $T^\ast(S^D)$ as the set of pairs $(x_I, J_{KL})$ in which $x_I$ is an unit vector in $\mathbb{R}^{D+1}$, $J_{KL}$ is a $(D+1)\times(D+1)$ skew-symmetric matrix, and $x_I$ and $J_{KL}$ satisfy
\begin{equation}
	J_{KL}=J_{KM}x^Mx_L-x_K J_{LM}x^M \label{eq:ConstraintJ}
\end{equation}
with momentum $p_I$ being defined by $p_J:=J_{IJ}x^I$. Based on this convention, the symplectic structure on $T^\ast(S^D)$ can be characterized by the Poisson bracket relations
\begin{equation}
\{J_{IJ},J_{KL}\}=-\delta_{IL}J_{JK}-\delta_{JK}J_{ IL}+\delta_{IK}J_{JL}+\delta_{JL }J_{IK},
\end{equation}
\begin{equation}
\{x_I,J_{JK}\}=\delta_{IK}x_J-\delta_{IJ}x_K,
\end{equation}
\begin{equation}
\{x_I,x_J\}=0.
\end{equation}

Let us now consider the quantum theory of the above constructions.
$J_{IJ}$ and $x_K$ should be replaced by self-adjoint operators $\hat{J}_{IJ}$ and $\hat{x}_K$ acting on the Hilbert space $L^2\left(S^D\right)$. The operators should satisfy $\hat{J}_{IJ}=-\hat{J}_{JI}$ and
\begin{equation}\label{[JJ]}
\frac{1}{\mathbf{i}\hbar}[\hat{J}_{IJ},\hat{J}_{KL}]=-\delta_{IL}\hat{J}_{JK}-\delta_{JK}\hat{J}_{ IL}+\delta_{IK}\hat{J}_{JL}+\delta_{JL }\hat{J}_{IK},
\end{equation}
\begin{equation}
\frac{1}{\mathbf{i}\hbar}[\hat{x}_I,\hat{J}_{JK}]=\delta_{IK}\hat{x}_J-\delta_{IJ}\hat{x}_K,
\end{equation}
\begin{equation}
[\hat{x}_I,\hat{x}_J]=0.
\end{equation}
We recognize this as a representation of the Euclidean Lie algebra e$(D+1)$=so$(D+1)\rtimes \mathbb{R}^{D+1}$, where the $\hat{J}_{IJ}$ represent the so$(D+1)$ sub-algebra according to Eq.\eqref{[JJ]}. The flux operators in $(D+1)$-dimensional LQG, typically denoted by $\hat{F}^{IJ}$, satisfy the same algebra (upto a constant) for suitable choices of surfaces and holonomies acted upon, see \cite{Bodendorfer:Qu} for details and the discussion in section \ref{sec:GeomOperators}.

It is important to implement the constraint \eqref{eq:ConstraintJ} also in the quantum theory. Otherwise, the $\hat{J}_{IJ}$ would have more degrees of freedom than the $p_I$ for $D>2$. As explained in \cite{hall2002coherent}, this restricts the allowed representations to (in our notation) simple ones, corresponding precisely to implementing the simplicity constraints enforcing $\pi^{aIJ} = 2 n^{[I} E^{a|J]}$ \cite{Bodendorfer:Ha}. Mathematically, these representations are realized as homogeneous harmonic functions on $S^D$ of degree $N$ denoted by $\mathfrak{H}_{D+1}^{N}$. In such a representation, the quadratic Casimir operator satisfies $ \hat J_{IJ} \hat J^{IJ} \propto N(N+D-1)$. We will discuss there representations in more detail in the next section.

\section{Perelomov coherent states for SO$(D+1)$}
The angular momentum operators can be represented on the space of square integrable functions on $S^D$ as
\begin{equation}
\hat{J}^{IJ}f(\bm{x})=-\mathbf{i}\hbar \left(x^I\frac{\partial}{\partial x^J}-x^J\frac{\partial}{\partial x^I}\right)f(\bm{x}), ~~~
\quad f(\bm{x})\in L^2\left( S^D\right).
\end{equation}
A comprehensive introduction of this representation space is given in \cite{vilenkin2013representation}. We will review the main points relevant for this paper.

The homogeneous harmonic functions of degree $N$ on the $D$-sphere ($S^D$) provide an irreducible representation space of SO$(D+1)$, denoted by $\mathfrak{H}_{D+1}^{N}$, and with dimensionality $\dim(\mathfrak{H}_{D+1}^{N})=\frac{(D+N-2)!(2N+D-1)}{(D-1)!N!}$. Introduce a subgroup series $SO(D+1)\supset SO(D)\supset SO(D-1)\supset ... \supset SO(2)_{\delta_1^{[I}\delta_2^{J]}}$ where $SO(2)_{\delta_1^{[I}\delta_2^{J]}}$ is the one-parameter subgroup of SO$(D+1)$ composed of rotations in the two-dimensional vector space spanned by $\{\delta_1^{I},\delta_2^{J}\}$. An orthogonal basis of the space $\mathfrak{H}_{D+1}^{N}$ can be given as $\{\Xi_{D+1}^{N,\mathbf{M}}(\bm{x})|\mathbf{M}:=M_1,M_2,...,M_{D-1}, N\geq M_1 \geq M_2\geq...\geq|M_{D-1}|, \bm{x}\in S^D\}$, or equivalently, in Dirac bracket notation as $\ket{N,\mathbf{M}}$ where $\mathbf{M}:=M_1,M_2,...,M_{D-1}$ with $N\geq M_1 \geq M_2\geq...\geq|M_{D-1}|$, and $N,M_1,... M_{D-2}\in\mathbb{N}$, $M_{D-1}\in \mathbb{Z}$. The labelling $N, \mathbf{M}$ of the function $\Xi_{D+1}^{N,\mathbf{M}}(\bm{x})$ can be interpreted as that $\Xi_{D+1}^{N,\mathbf{M}}(\bm{x})$ belongs to the series of space $\mathfrak{H}_{2}^{M_{D-1}}\subset \mathfrak{H}_{3}^{M_{D-2}}\subset...\subset\mathfrak{H}_{D}^{M_{1}}\subset\mathfrak{H}_{D+1}^{N}$ which are the irreducible representation spaces labeled by $M_{D-1},...,M_2,M_1, N$ of the series of groups $SO(2)_{\delta_1^{[I}\delta_2^{J]}} \subset SO(3)\subset ... \subset SO(D)\subset SO(D+1)$ respectively \cite{vilenkin2013representation}. Based on this convention, the corresponding inner product is given by
\begin{equation}
\langle N,\mathbf{M}|N,\mathbf{M}'\rangle:=\int_{S^D} d\bm{x} \, \overline{\Xi_{D+1}^{N,\mathbf{M}}(\bm{x})}\Xi_{D+1}^{N,\mathbf{M}'}(\bm{x})=\delta_{\mathbf{M},\mathbf{M}'}
\end{equation}
with $\delta_{\mathbf{M},\mathbf{M}'}=1$ if $\mathbf{M}=\mathbf{M}'$ and zero otherwise,
where $d\bm{x}$ is the normalized invariant measure on $S^D$. Also, an element $g\in$ SO$(D+1)$ act on a spherical harmonic function on D-sphere $f(\bm{x})$ as
  \begin{equation}\label{gfx}
  g\circ f(\bm{x})=f(g^{-1}\circ\bm{x}).
  \end{equation}
  The general form of the functions $\Xi_{D+1}^{N,\mathbf{M}}(\bm{x})$ is not needed for this paper, several special examples are provided below.

Let us introduce the basis $\{X^{IJ}\}$ of so$(D+1)$, it is given by $(X^{IJ})_{\text{def.}}:=2\delta^I_{[K}\delta^J_{L]}$ in the definition representation space of SO$(D+1)$ and it acts on the spherical harmonic function as
 \begin{equation}
 X^{IJ}f(\bm{x}):=\frac{d}{dt}f(e^{-tX^{IJ}}\circ\bm{x})|_{t=0}=(x^I\frac{\partial}{\partial x^J}-x^J\frac{\partial}{\partial x^I})f(\bm{x}).
 \end{equation}
 Then, the Cartan subalgebra $\mathcal{C}$ of so$(D+1)$ can be generated by $C_{\tilde{k}} = \mathbf{i}X_{2\tilde{k}-1,2\tilde{k}}$, $\tilde{k} = 1, . . . ,[\frac{D+1}{2}]$, and we denote by $\mathbf{e}_{\tilde{k}}$ the generators of the dual of $\mathcal{C}$, $\mathbf{e}_{\tilde{k}}(C_{\tilde{j}})=\delta_{\tilde{k}\tilde{j}}$. Now, the highest weight vector of the representation space is given by $N\mathbf{e}_1$, and the special state which corresponds to the highest weight vector $N\mathbf{e}_1$ is denoted by $|N\mathbf{e}_1\rangle=|N,\delta_1^{[I}\delta_2^{J]}\rangle:=|N, \mathbf{M}=(N,...,N)\rangle$, which can also be expressed as the homogeneous harmonic function
 \begin{equation}\label{N1I2J}
 \Xi_{D+1}^{N,\delta_1^{[I}\delta_2^{J]}}(\bm{x}):=\text{c}_N\frac{(\bm{x}\cdot\bm{\delta}_1 +\mathbf{i}\bm{x}\cdot\bm{\delta}_2)^N}{r^2}=\text{c}_N(x_1 +\mathbf{i}x_2)^N,
 \end{equation}
 where $r^2=\bm{x}\cdot\bm{x}=1$ and $\text{c}_N$ is the normalization factor given by
\begin{equation}
\text{c}_N=\frac{1}{\sqrt{2\pi}}\prod_{d=2}^D \left( \frac{2^N\Gamma(N+\frac{d-1}{2})}{\Gamma(\frac{d-1}{2})} \sqrt{\frac{(d-2)!(2N+d-1)}{(2N+d-2)!(d-1)}} \right).
\end{equation}
Also, by introducing the spherical coordinate system $(\xi_{D},...,\xi_2,\xi_1)$ on $S^D$ which links to $(x_1,...,x_{D+1})$ by
\begin{eqnarray}
x_{D+1}&=&\cos\xi_D,\\\nonumber
x_D&=&\sin\xi_D\cos\xi_{D-1},\\\nonumber
x_{D-1}&=&\sin\xi_D\sin\xi_{D-1}\cos\xi_{D-2},\\\nonumber
&&...\\\nonumber
x_2&=&\sin\xi_D\sin\xi_{D-1}...\sin\xi_{2}\sin\xi_1,\\\nonumber
x_1&=&\sin\xi_D\sin\xi_{D-1}...\sin\xi_{2}\cos\xi_1
\end{eqnarray}
with $0\leq\xi_D,\xi_{D-1},...,\xi_{2}\leq\pi$, $0\leq\xi_1<2\pi$,
the function $\Xi_{D+1}^{N,\delta_1^{[I}\delta_2^{J]}}(\bm{x})$ can be re-expressed as
\begin{equation}
\Xi_{D+1}^{N,\delta_1^{[I}\delta_2^{J]}}(\bm{x})=\Xi_{D+1}^{N,\delta_1^{[I}\delta_2^{J]}}(\bm{\xi}) =\textrm{c}_N\sin^N\xi_D\sin^N\xi_{D-1}...\sin^N\xi_{2}e^{\mathbf{i}N\xi_1}.
\end{equation}

Following the construction procedure of the Perelomov coherent states introduced in \cite{GeneralizedCoherentStates},  we can construct the SO$(D+1)$ Perelomov coherent states in the simple representation space based on the state $|N\mathbf{e}_1\rangle=|N,\delta^{[I}_1\delta^{J]}_2\rangle$ which corresponds to the highest weight vector. The result is the system of states $\{|N, g\rangle\}$, $|N, g\rangle:=g|N\mathbf{e}_1\rangle=g|N,V_0\rangle$ with $V_0:=\delta^{[I}_1\delta^{J]}_2$, where $g$ are elements of the group SO$(D+1)$ which acts on $|N\mathbf{e}_1\rangle$ following Eq.\eqref{gfx}. More explicitly, a coherent state $|N, g\rangle$ is determined by a point $V=V(g):=gV_0g^{-1}$ in the coset space $Q_{D-1}:=SO(D+1)/(SO(2)\times SO(D-1))$, where $SO(2)\times SO(D-1)$ is the maximal isotropic subgroup of $V_0$. Notice that we can decompose $g$ as $g=u\bar{u}e^{(\alpha V^{IJ}_0X_{IJ})}$ with $u\in Q_{D-1}, \bar{u}\in SO(D-1)$, $0\leq\alpha<2\pi$ and $e^{(\alpha V^{IJ}_0X_{IJ})}\in SO(2)$. Hence, we can give another formulation $|N, V\rangle$ of SO$(D+1)$ Perelomov coherent states by the relation $|N, g\rangle=\exp(\mathbf{i}N\alpha)|N, V\rangle$ with $V=gV_0g^{-1}$, which can be proved as follows. Notice that $|N,V_0\rangle$ corresponds to the function \eqref{N1I2J} so that $|N, g\rangle=g|N,V_0\rangle$ corresponds to the function
 \begin{equation}
  g\circ\Xi_{D+1}^{N,V_0}(\bm{x})=\Xi_{D+1}^{N,V_0}(g^{-1}\circ\bm{x})=\text{c}_N\frac{(\bm{x}\cdot(g\circ\bm{\delta}_1) +\mathbf{i}\bm{x}\cdot(g\circ\bm{\delta}_2))^N}{r^2}=\exp(\mathbf{i}N\alpha)\cdot \Xi_{D+1}^{N,gV_0g^{-1}}(\bm{x}).
 \end{equation}
 Then we finish the proof of that $|N, g\rangle=\exp(\mathbf{i}N\alpha)|N, V\rangle$ with $V=gV_0g^{-1}$. Now, let us begin to discuss the general properties of these Perelomov coherent states as follows.

\begin{enumerate}
\item \textit{The homogeneous harmonic function $\Xi^{N,V}_{D=1}(\bm{x})$ on $S^D$ corresponding to the Perelomov coherent states $|N,V\rangle$ can be regarded as wave functions of a particle moving on $S^D$, and the probability amplitude given by the these wave functions is
    \begin{equation}
    |\Xi^{N,V}_{D=1}(\bm{x})|^2=\textrm{c}_N^2(x_1^2+x_2^2)^N=c_N^2\sin^{2N}\xi_D\sin^{2N}\xi_{D-1}...\sin^{2N}\xi_{2},
    \end{equation}
    which is peaked at the 1-dimensional circle labelled by $\xi_{D}=\xi_{D-1}=...=\xi_2=\frac{\pi}{2}$ or $x_3=x_4=...=x_{D+1}=0$ in $S^D$ in the large $N$ limit.}
  \item \textit{The angular momentum operators sandwiched between coherent states satisfy $\langle N,V|\hat{J}^{IJ}|N, V\rangle=2N\hbar V^{IJ}$, and their uncertainties read
  \begin{equation} \bigtriangleup\vev{\hat{J}^{IJ}} :=\sqrt{\left|\sum_{I,J}\vev{\hat{J}^{IJ}}\vev{\hat{J}^{IJ}}-\sum_{I,J}\vev{\hat{J}^{IJ}\hat{J}^{IJ}}\right|} =\sqrt{2N(D-1)}\hbar,
  \end{equation}
 which tends to zero in the limit $N\hbar\rightarrow1$, $\hbar\rightarrow0$.}

      \textbf{Proof.}  Without loss of generality, we choose $|N, V\rangle$ as $|N\mathbf{e}_1\rangle$ and find
      \begin{align}\label{Jf}
\hat{J}_{12}|N\mathbf{e}_1\rangle&=N\hbar|N\mathbf{e}_1\rangle, \hspace{-3cm} &&\\
\hat{J}_{IJ}|N\mathbf{e}_1\rangle&=0, && I,J\neq1,2,\\
\langle N\mathbf{e}_1|\hat{J}_{IJ}|N\mathbf{e}_1\rangle&=0, && I=1\ \text{or} \ 2,J\neq1,2,\\
\langle N\mathbf{e}_1|\hat{J}_{IJ}\hat{J}_{IJ}|N\mathbf{e}_1\rangle &=\frac{N}{2}\hbar^2, && I=1\ \text{or} \ 2,J\neq1,2,
\end{align}
and
\begin{equation}\label{Jl}
\Delta \vev{\hat{J}_{IJ}} :=\sqrt{\vev{\hat{J}_{IJ}\hat{J}_{IJ}}-\left(\vev{\hat{J}_{IJ}}\right)^2}=\sqrt{\frac{N}{2}}\hbar,\qquad I=1\ \text{or} \ 2,J\neq1,2,
\end{equation}
where we used the shorthand $\vev{\ldots}\equiv\langle N\mathbf{e}_1|\ldots|N\mathbf{e}_1\rangle$. The equations above about the expectation values can be summarized as
\begin{equation}
\vev{\hat{J}_{IJ}}:=\langle N\mathbf{e}_1|\hat{J}_{IJ}|N\mathbf{e}_1\rangle=2N\hbar\delta^1_{[I}\delta^2_{J]}.
\end{equation}
Further, the rest of the equations imply that the state $|N\mathbf{e}_1\rangle$ minimizes the uncertainty
\begin{eqnarray}
\Delta \left(\vev{\hat{J}_{IJ}} \right)&:=&\sqrt{\sum_{I,J}\langle N\mathbf{e}_1|\hat{J}_{IJ}\hat{J}^{IJ}|N\mathbf{e}_1\rangle-\sum_{I,J}\langle N\mathbf{e}_1|\hat{J}_{IJ}|N\mathbf{e}_1\rangle\langle N\mathbf{e}_1|\hat{J}^{IJ}|N\mathbf{e}_1\rangle}\\\nonumber
&=&\sqrt{2N(N+D-1)-2N^2}\hbar=\sqrt{2N(D-1)}\hbar,
\end{eqnarray}
which tends to zero in the limit $N\hbar\rightarrow1$, $\hbar\rightarrow0$. This result can be extended to state $|N,V\rangle$ immediately based on the definition $|N,V\rangle=e^{-\mathbf{i}N\alpha}g|N\mathbf{e}_1\rangle, V=gV_0g^{-1}$. This finishes our proof.
$\square$

  \item \textit{The coherent states minimize the Heisenberg uncertainty relation of angular momentum operators $\hat{J}_{IJ}$: the inequality
      \begin{equation}
      \left(\bigtriangleup\!\vev{\hat{J}_{IJ}}\right)^2 \left(\bigtriangleup\!\vev{\hat{J}_{KL}}\right)^2\ \geq\ \frac{1}{4}\left|\vev{[\hat{J}_{IJ},\hat{J}_{KL}]}\right|^2
      \end{equation}
      is saturated for the state $|N, g\rangle$}.

      \textbf{Proof.} First, let us prove it for state $|N\mathbf{e}_1\rangle$. Based on the Eqs.(\ref{Jf})-(\ref{Jl}), and the relation $[\hat{J}_{IJ},\hat{J}_{KL}]=\mathbf{i}\hbar(-\delta_{IL}\hat{J}_{JK}-\delta_{JK}\hat{J}_{IL} +\delta_{IK}\hat{J}_{JL}+\delta_{JL}\hat{J}_{IK})$, it is easy to see that\\ \mbox{$\left(\bigtriangleup\!\vev{\hat{J}_{IJ}}\right)^2 \left(\bigtriangleup\!\vev{\hat{J}_{KL}}\right)^2\ =\ \frac{1}{4}\left|\vev{[\hat{J}_{IJ},\hat{J}_{KL}]}\right|^2=0$} holds except in the case where $[\hat{J}_{IJ},\hat{J}_{KL}]$ contains a term proportion to $\hat{J}_{12}$. In this case, we always have\\
       $\left(\bigtriangleup\!\vev{\hat{J}_{IJ}}\right)^2 \left(\bigtriangleup\!\vev{\hat{J}_{KL}}\right)^2\ =\ \frac{1}{4}\left|\vev{[\hat{J}_{IJ},\hat{J}_{KL}]}\right|^2=\frac{N^2}{4}\hbar^4$. Now let us extend the result to general coherent states. For the transformed angular momentum operator components $\tilde{\hat{J}}_{IJ}:=g_V\hat{J}_{IJ}g_V^{-1}$, the state $|N\mathbf{e}_1\rangle$ also minimizes the uncertainty relation $\left(\bigtriangleup\!\vev{\hat{J}_{IJ}}\right)^2 \left(\bigtriangleup\!\vev{\hat{J}_{KL}}\right)^2\ \geq\ \frac{1}{4}\left|\vev{[\hat{J}_{IJ},\hat{J}_{KL}]}\right|^2$. Then, it is easy to see that the relation is minimized for the state $|N, V\rangle$ from its definition. \ $\square$
  \item \textit{The system of coherent states $\{|N, g\rangle\}$ gives a complete basis of $\mathfrak{H}_{D+1}^N$, and the resolution of unit can be written as
\begin{equation}\label{iden1}
  \dim\left(\mathfrak{H}_{D+1}^N\right)\int_{Q_{D-1}}dV|N,V\rangle\langle N,V|=\mathbb{I}_{\mathfrak{H}_{D+1}^N},
\end{equation}
where $\int_{Q_{D-1}}dV=1$, $dV$ is the invariant measure induce by the Haar measure of SO$(D+1)$.}\\
      \textbf{Proof.}   Let us consider the operator $\hat{B}:=\int_{Q_{D-1}}dV|N,V\rangle\langle N,V|$. Due to the invariance of the measure $dV$, one has at once $g\hat{B}g^{-1}=\hat{B}$. Thus $\hat{B}$ commutes with all group elements $g$ and must be equal to the identity $\mathbb{I}_{\mathfrak{H}_{D+1}^N}$ in $\mathfrak{H}_{D+1}^N$ times a numerical factor (the representation space $\mathfrak{H}_{D+1}^N$ is irreducible). To fix the numerical factor, it is useful to calculate the trace of $\hat{B}$, which gives
      \begin{equation}
      \text{tr}(\hat{B})=\text{tr}\left(\int_{Q_{D-1}}dVg(V)|N\mathbf{e}_1\rangle\langle N\mathbf{e}_1|(g(V))^{-1}\right)=\text{tr}\left(\int_{Q_{D-1}}dV|N\mathbf{e}_1\rangle\langle N\mathbf{e}_1|\right)=1,
      \end{equation}
      Comparing with $\text{tr}(\mathbb{I}_{\mathfrak{H}_{D+1}^N})=\dim\left(\mathfrak{H}_{D+1}^N\right)$, we immediately get\\ \mbox{$\dim\left(\mathfrak{H}_{D+1}^N\right)\int_{Q_{D-1}}dV|N,V\rangle\langle N,V|=\mathbb{I}_{\mathfrak{H}_{D+1}^N}$}.   \ $\square$
  \item \textit{The coherent states $|N, V\rangle$ and $|N, V'\rangle$ are not mutually orthogonal unless $[V^{IJ}X_{IJ},V'^{KL}X_{KL}]=0$ and $V^{IJ}\neq V'^{KL}$}.

      \textbf{Proof.} Generally, a Perelomov coherent state of SO$(D+1)$ in a simple representation space labelled by $N$ is given by $|N, V\rangle$, where $V=V^{IJ}=m^{[I}n^{J]}$ is the labelling bi-vector of the state $|N, V\rangle$ and $m^I$, $n^I$ are unit vectors in $\mathbb{R}^{D+1}$. The labelling bi-vector has the property that $V^{IJ}X_{IJ}|N, V\rangle=\mathbf{i}N|N, V\rangle$ and the operator $X^{IJ}$  is peaked at $2\mathbf{i}NV^{IJ}$ with relative uncertainty $\sim \frac{1}{\sqrt{N}}$ (see \cite{long2019coherent}).
      We now turn to the inner product of these coherent states. Without loss of generality, we can fix $V'=\delta_1^{[I}\delta_2^{J]}$, and define a projection $\eta^I_J:=(\delta_1)^I(\delta_1)_J+(\delta_2)^I(\delta_2)_J$ which projects a vector to the $2$-dimensional vector space spanned by $\delta_1^I$ and $\delta_2^J$, and also its complement $\bar{\eta}^I_J:=\delta^I_J-\eta^I_J$. Now, for $V=V^{IJ}=m^{[I}n^{J]}$, we differentiate three cases: (i), $\bar{\eta}_I^K\bar{\eta}_J^Lm^{[I}n^{J]}=0$;  (ii), $\bar{\eta}_I^K\bar{\eta}_J^Lm^{[I}n^{J]}\neq0, \eta_I^K\eta_J^Lm^{[I}n^{J]}=0$; (iii), $\bar{\eta}_I^K\bar{\eta}_J^Lm^{[I}n^{J]}\neq0, \eta_I^K\eta_J^Lm^{[I}n^{J]}\neq0$. In the following, we will discuss each one seperately.

\textbf{Case (i)}: In this case, the labelling bi-vector $V=V^{IJ}=m^{[I}n^{J]}$ of the coherent state $|N, V\rangle$ can be re-expressed as $V^{IJ}=v_1^{[I}v_2^{J]}$ where $v_1^I, v_2^J$ are unit vectors satisfying $\eta_I^Jv_1^I=v_1^J$ and $v_1^Iv_2^J\delta_{IJ}=0$. We define $\cos\theta=|\eta^I_Jv_2^J|$. From a result in \cite{long2019coherent}, it follows that $\langle N, V|N, V'\rangle=e^{\mathbf{i}N\phi}(\frac{1+\cos\theta}{2})^N$ where $e^{\mathbf{i}N\phi}$ is the phase factor of the Perelomov coherent states.

\textbf{Case (ii)}: In this case, the labelling bi-vector $V=V^{IJ}=m^{[I}n^{J]}$ of the coherent state $|N, V\rangle$ can be re-expressed as $V^{IJ}=v_1^{[I}v_2^{J]}$, where $v_2^I\delta_1^J\delta_{IJ}=0$, $v_2^I\delta_2^J\delta_{IJ}=0$ and $\cos\theta=|\eta^I_Jv_1^J|$. Let us decompose $v_1^I$ as $v_1^I=w^I+w'^I$ where $\eta^I_Jv_1^J=w^I$ and $\eta^I_Jw'^J=0$, and denote these vectors with indeces $I,J,K,..$ by $\bm{\delta}_1, \bm{\delta}_1, \bm{v}_1, \bm{v}_2,\bm{w}, \bm{w}'$, then we have $|\bm{w}|=\cos\theta$. Based on these definitions, the coherent states $|N, V\rangle$ and $|N, V'\rangle$ can be expressed as a homogeneous harmonic function $\Xi^{N,V}_{D+1}(\bm{x}):=\text{c}_N(\bm{x}\cdot \bm{v}_1+\mathbf{i}\bm{x}\cdot \bm{v}_2)^N$ and $\Xi^{N,V'}_{D+1}(\bm{x}):=\text{c}_N(\bm{x}\cdot \bm{\delta}_1+\mathbf{i}\bm{x}\cdot \bm{\delta}_2)^N$ respectively. Let us introduce a subgroup series $SO(D+1)\supset SO(2)_{V'}\times SO(D-1)\supset SO(2)_{V'}\times SO(D-2)\supset... \supset SO(2)_{V'}\times SO(2)$ where $SO(2)_{V'}$ gives the rotation in the 2-dimensional vector space spanned by $\{\delta_1^I, \delta_2^I\}$. Based on this series, we can decompose the space $\mathfrak{H}^N_{D+1}$ of homogeneous harmonic $D$-spherical function with degree $N$ as \cite{ vilenkin2013representation}
\begin{equation}
\mathfrak{H}^N_{D+1}=\bigoplus_{P,Q}\left(\mathfrak{H}^P_{2}\otimes \mathfrak{H}^Q_{D-1}\right),\quad P+Q+2O=N,~~  O=0,1,..,\left[\frac{N}{2}\right], \label{eq:HND1}
\end{equation}
where $\mathfrak{H}^P_{2}$ and $\mathfrak{H}^Q_{D-1}$ are homogeneous harmonic functions with degree $P$ and $Q$ on the $1$-sphere and $(D-2)$-sphere respectively. Now, following the discussion in \cite{vilenkin2013representation}, we know that $\Xi^{N,V'}_{D+1}(\bm{x})\in (\mathfrak{H}^N_{2}\otimes \mathfrak{H}^0_{D-1})\subset \mathfrak{H}^N_{D+1}$, and conclude that only the projection of $\Xi^{N,V}_{D+1}(\bm{x})$ into $(\mathfrak{H}^N_{2}\otimes \mathfrak{H}^0_{D-1})$ will contribute to the inner product $\langle N, V|N, V'\rangle$. Let us write $\Xi^{N,V}_{D+1}(\bm{x}):=\text{c}_N(\bm{x}\cdot \bm{v}_1+\mathbf{i}\bm{x}\cdot \bm{v}_2)^N$ as
\begin{eqnarray}
\Xi^{N,V}_{D+1}(\bm{x})&:=&\text{c}_N(\bm{x}\cdot \bm{v}_1+\mathbf{i}\bm{x}\cdot \bm{v}_2)^N\\\nonumber
&=&\text{c}_N(\bm{x}\cdot(\bm{w}+\bm{w}')+\mathbf{i}\bm{x}\cdot \bm{v}_2)^N\\\nonumber
&=&\text{c}_N\sum_{N'=0}^N\frac{N!}{N'!(N-N')!}(\bm{x}\cdot\bm{w})^{N'}(\bm{x}\cdot\bm{w}'+\mathbf{i}\bm{x}\cdot \bm{v}_2)^{(N-N')}.
\end{eqnarray}
It is easy to see that the projection of $\Xi^{N,V}_{D+1}(\bm{x})$ into $(\mathfrak{H}^N_{2}\otimes \mathfrak{H}^0_{D-1})$ is given by the term with $N'=N$ in the above sum, that is
 \begin{eqnarray}
 \tilde{\Xi}^{N,V}_{D+1}(\bm{x})&:=&\text{c}_N(\bm{x}\cdot\bm{w})^{N} =\frac{\text{c}_N\cos^N\theta}{2^N}\left((\bm{x}\cdot\bar{\bm{w}} +\mathbf{i}\bm{x}\cdot\bar{\bm{w}}')+(\bm{x}\cdot\bar{\bm{w}} -\mathbf{i}\bm{x}\cdot\bar{\bm{w}}')\right)^N\\\nonumber
 &=&\frac{\text{c}_N\cos^N\theta}{2^N} \sum_{N''=0}^N\frac{N!}{N''!(N-N'')!}(\bm{x}\cdot\bar{\bm{w}} +\mathbf{i}\bm{x}\cdot\bar{\bm{w}}')^{N''}(\bm{x}\cdot\bar{\bm{w}} -\mathbf{i}\bm{x}\cdot\bar{\bm{w}}')^{(N-N'')},
 \end{eqnarray}
 wherein $\cos\theta=|\bm{w}|$, $\bar{\bm{w}}:=\bm{w}/|\bm{w}|$, and $\bar{w}'^{I}$ is a unit vector defined by $\bar{w}^{[I}\bar{w}'^{J]}=\delta_1^{[I}\delta_2^{J]}$. Now, we can calculate that
\begin{eqnarray}
\langle N, V|N, V'\rangle&=&\int_{S^D}d\bm{x} \, \overline{\tilde{\Xi}^{N,V}_{D+1}(\bm{x})}\Xi^{N,V'}_{D+1}(\bm{x})\\\nonumber
&=&\int_{S^D}d\bm{x} \,\frac{\cos^N\theta}{2^N}\overline{\text{c}_N (\bm{x}\cdot\bar{\bm{w}} +\mathbf{i}\bm{x}\cdot\bar{\bm{w}}')^{N}}\Xi^{N,V'}_{D+1}(\bm{x})\\\nonumber
&=&\frac{\cos^N\theta}{2^N}e^{\mathbf{i}N\phi}.
\end{eqnarray}

\textbf{Case (iii)}: In this case, the labelling bi-vector $V=V^{IJ}=m^{[I}n^{J]}$ of the coherent state $|N, V\rangle$ can be re-expressed as $V^{IJ}=v_1^{[I}v_2^{J]}$ where $v_1^J\eta_J^I\neq0, v_1^J\bar{\eta}_J^I\neq0, v_2^J\eta_J^I\neq0, v_2^J\bar{\eta}_J^I\neq0$, and $\cos\theta_1=|\eta^I_Jv_1^J|$, $\cos\theta_2=|\eta^I_Jv_2^J|$. Let us decompose $v_1^I$ as $v_1^I=s_1^I+s_1'^I$ and $v_2^I$ as $v_2^I=s_2^I+s_2'^I$, where $\eta^I_Js_1^J=s_1^I$, $\eta^I_Js_2^J=s_2^I$ and $\eta^I_Js_2'^J=0$, $\eta^I_Js_2'^J=0$. Similarly, we omit the indices $I,J,...$ and use bold font to represent vectors, and express the coherent states $|N, V\rangle$ and $|N, V'\rangle$ as homogeneous harmonic functions $\Xi^{N,V}_{D+1}(\bm{x}):=\text{c}_N(\bm{x}\cdot \bm{v}_1+\mathbf{i}\bm{x}\cdot \bm{v}_2)^N$ and $\Xi^{N,V'}_{D+1}(\bm{x}):=\text{c}_N(\bm{x}\cdot \bm{\delta}_1+\mathbf{i}\bm{x}\cdot \bm{\delta}_2)^N$ respectively. Considering the same decomposition of $\mathfrak{H}^N_{D+1}$ as in \eqref{eq:HND1}, we again get the result that only the projection of $\Xi^{N,V}_{D+1}(\bm{x})$ into $(\mathfrak{H}^N_{2}\otimes \mathfrak{H}^0_{D-1})$ will contribute to the inner product $\langle N, V|N, V'\rangle$. Let us expand $\Xi^{N,V}_{D+1}(\bm{x}):=\text{c}_N(\bm{x}\cdot \bm{v}_1+\mathbf{i}\bm{x}\cdot \bm{v}_2)^N$ as
\begin{eqnarray}
\Xi^{N,V}_{D+1}(\bm{x})&:=&\text{c}_N(\bm{x}\cdot \bm{v}_1+\mathbf{i}\bm{x}\cdot \bm{v}_2)^N\\\nonumber
&=&\text{c}_N(\bm{x}\cdot(\bm{s}_1+\bm{s}_1')+\mathbf{i}\bm{x}\cdot (\bm{s}_2+\bm{s}'_2))^N\\\nonumber
&=&\text{c}_N\sum_{N'=0}^N\frac{N!}{N'!(N-N')!}(\bm{x}\cdot\bm{s}_1+\mathbf{i} \bm{x}\cdot\bm{s}_2)^{N'}(\bm{x}\cdot\bm{s}'_1+\mathbf{i}\bm{x}\cdot \bm{s}'_2)^{(N-N')}.
\end{eqnarray}
It is easy to see that only the term $ \tilde{\Xi}^{N,V}_{D+1}(\bm{x})$ in the decomposition of $\Xi^{N,V}_{D+1}(\bm{x})$ projecting into $(\mathfrak{H}^N_{2}\otimes \mathfrak{H}^0_{D-1})$ will not vanish, given by
\begin{eqnarray}
 && \tilde{\Xi}^{N,V}_{D+1}(\bm{x}) \\\nonumber
 &=& \text{c}_N(\bm{x}\cdot\bm{s}_1+\mathbf{i} \bm{x}\cdot\bm{s}_2)^{N} =  \text{c}_N(\bm{x}\cdot\bar{\bm{s}}_1\cos\theta_1+\mathbf{i} \bm{x}\cdot\bar{\bm{s}}_2\cos\theta_2)^{N} \\\nonumber
   &=&\text{c}_N((\bm{x}\cdot\bar{\bm{s}}_1 +\mathbf{i}\bm{x}\cdot\check{\bar{\bm{s}}}_1)\frac{\cos\theta_1}{2}+(\bm{x}\cdot\bar{\bm{s}}_1 -\mathbf{i}\bm{x}\cdot\check{\bar{\bm{s}}}_1)\frac{\cos\theta_1}{2} \\\nonumber &&+(\bm{x}\cdot\check{\bar{\bm{s}}}_2+\mathbf{i} \bm{x}\cdot\bar{\bm{s}}_2)\frac{\cos\theta_2}{2}-(\bm{x}\cdot\check{\bar{\bm{s}}}_2-\mathbf{i} \bm{x}\cdot\bar{\bm{s}}_2)\frac{\cos\theta_2}{2})^{N} \\\nonumber
   &=&\text{c}_N\left((\bm{x}\cdot\bar{\bm{s}}_1 +\mathbf{i}\bm{x}\cdot\check{\bar{\bm{s}}}_1)(\frac{\cos\theta_1}{2}+ e^{\mathbf{i}\gamma_{(\bar{\bm{s}}_1,\bar{\bm{s}}_2)}}\frac{\cos\theta_2}{2})+(\bm{x}\cdot\bar{\bm{s}}_1 -\mathbf{i}\bm{x}\cdot\check{\bar{\bm{s}}}_1)(\frac{\cos\theta_1}{2} -e^{-\mathbf{i}\gamma_{(\bar{\bm{s}}_1,\bar{\bm{s}}_2)}}\frac{\cos\theta_2}{2}) \right)^{N},
   \end{eqnarray}
where $\bar{\bm{s}}_1$ and $\bar{\bm{s}}_2$ are unit vectors defined by $\cos\theta_1\cdot\bar{\bm{s}}_1=\bm{s}_1$ and $\cos\theta_2\cdot\bar{\bm{s}}_2=\bm{s}_2$ respectively, $\check{\bar{\bm{s}}}_1$ and $\check{\bar{\bm{s}}}_2$ are unit vectors defined by $\bar{s}^{[I}_1\check{\bar{s}}^{J]}_1=\check{\bar{s}}^{[I}_2\bar{s}^{J]}_2=\delta^{[I}_1\delta_2^{J]}$, and $\gamma_{(\bar{\bm{s}}_1,\bar{\bm{s}}_2)}$ is the angle defined by $\exp(\gamma_{(\bar{\bm{s}}_1,\bar{\bm{s}}_2)}\delta^{[I}_1\delta_2^{J]}\tau_{IJ})\cdot\bar{\bm{s}}_1 =\check{\bar{\bm{s}}}_2$ with $\delta^{[I}_1\delta_2^{J]}\tau_{IJ}$ being the generator of the rotation which rotates $\delta^{I}_1$ to $\delta^{I}_2$ by the angle $\frac{\pi}{2}$.
Now, we can calculate that
\begin{eqnarray}
\langle N, V|N, V'\rangle&=&\int_{S^D}d\bm{x} \overline{\tilde{\Xi}^{N,V}_{D+1}(\bm{x})}\Xi^{N,V'}_{D+1}(\bm{x})\\\nonumber
&=&\int_{S^D}d\bm{x}\overline{\text{c}_N\left((\bm{x}\cdot\bar{\bm{s}}_1 +\mathbf{i}\bm{x}\cdot\check{\bar{\bm{s}}}_1)\left(\frac{\cos\theta_1}{2}+ e^{\mathbf{i}\gamma_{\left(\bar{\bm{s}}_1,\bar{\bm{s}}_2\right)}}\frac{\cos\theta_2}{2}\right)\right)^{N}}\Xi^{N,V'}_{D+1}(\bm{x})\\\nonumber
&=&\left(\frac{\cos\theta_1}{2}+ e^{\mathbf{i}\gamma_{(\bar{\bm{s}}_1,\bar{\bm{s}}_2)}}\frac{\cos\theta_2}{2}\right)^{N}e^{\mathbf{i}N\phi}.
\end{eqnarray}

Generally, we can also regard the case (i) as a special case of case (iii) with $\theta_1=0$, $\gamma_{(\bar{\bm{s}}_1,\bar{\bm{s}}_2)}=0$, and case (ii) as special cases of case (iii) with $\theta_2=\frac{\pi}{2}$.
We conclude that $\langle N, V|N, V'\rangle=0$ only when $\theta_1=\theta_2=\frac{\pi}{2}$, which is equivalent to require that $[V^{IJ}X_{IJ},V'^{KL}X_{KL}]=0$.  This finishes our proof. $\square$

A special property of the angle $\gamma_{(\bar{\bm{s}}_1,\bar{\bm{s}}_2)}$ is worth to be discussed. Recall that $\bm{v}_1=\cos\theta_1\bar{\bm{s}}_1+\sin\theta_1\tilde{\bm{s}}_1$ and $\bm{v}_2=\cos\theta_2\bar{\bm{s}}_2+\sin\theta_2\tilde{\bm{s}}_2 =\cos\theta_2(\sin\gamma_{(\bar{\bm{s}}_1,\bar{\bm{s}}_2)}\bar{\bm{s}}_1 \pm\cos\gamma_{(\bar{\bm{s}}_1,\bar{\bm{s}}_2)}\check{\bar{\bm{s}}}_1)+\sin\theta_2\tilde{\bm{s}}_2$, where $\tilde{\bm{s}}_1$, $\tilde{\bm{s}}_2$, $\bar{\bm{s}}_1$ and $\bar{\bm{s}}_2$ are unit vectors, $\cos\theta_1\bar{\bm{s}}_1$ and $\cos\theta_2\bar{\bm{s}}_2$ are the projections of $\bm{v}_1$ and $\bm{v}_2$ into the 2-plane spanned by $\{\delta_1^I,\delta_2^I\}$ respectively. Notice that due to $\bm{v}_1\cdot\bm{v}_2=0$, we can immediately get
\begin{equation}\label{gamma}
\sin\gamma_{(\bar{\bm{s}}_1,\bar{\bm{s}}_2)} =-\tan\theta_1\tan\theta_2\tilde{\bm{s}}_1\cdot\tilde{\bm{s}}_2,\quad \tilde{\bm{s}}_1\cdot\tilde{\bm{s}}_2\leq1.
\end{equation}
From now on, the set $(\theta_1,\theta_2,\gamma=\gamma_{(\bar{\bm{s}}_1,\bar{\bm{s}}_2)})$ introduced in above \textbf{Case (iii)} will be called the set of angles between the bi-vectors $V$ and $V'$.

      \item \textit{The matrix element function $\Xi_{D+1}^{N,V,V'}(g):=\sqrt{\dim(\mathfrak{H}_{D+1}^{N})}\langle N,V'|g|N,V \rangle$, $g\in SO(D+1)$ is sharply peaked at the subgroup $SO(D+1)_{(V,V')}$ of SO$(D+1)$ in the large $N$ limit, where $SO(D+1)_{(V,V')}$ is composed of all elements $g\in SO(D+1)$ which satisfy $gVg^{-1}=V'$.}

          This property is obvious from the calculation of the inner product of $\langle N,V'|N,V \rangle$ in the proof of the last item. It is easy to see that all the elements of $SO(D+1)_{(V,V')}$ can be reproduced by $g_{VV'}g_V$ when $g_V\in (SO(2)\times SO(D-1))_V$ runs all over $ (SO(2)\times SO(D-1))_V$, where $g_{VV'}$ is an arbitrary but fixed element of $SO(D+1)_{(V,V')}$, and $(SO(2)\times SO(D-1))_V\subset SO(D+1)$ which is the maximal subgroup of SO$(D+1)$ which fixes $V$. A special case of the matrix element function $\Xi_{D+1}^{N,V,V'}(g)$ is $V=V'$, which is peaked at the subgroup $(SO(2)\times SO(D-1))_V$. Further, we can fix $g$ as the identity of SO$(D+1)$ to obtain the functions $\Xi_{D+1}^{N,V'}(V):=\sqrt{\dim(\mathfrak{H}_{D+1}^{N})}\langle N,V'|N,V \rangle$ on $Q_{D-1}$. For similar reasons, we can also conclude that $\Xi_{D+1}^{N,V'}(V)$ is sharply peaked at $V=V'$, which can be represented as
           \begin{align}\label{PV}
           \lim_{N\rightarrow\infty}\left|\Xi_{D+1}^{N,V'}(V)\right|^2\left|_{V=V'}\right.&= \lim_{N\rightarrow\infty}\dim\left(\mathfrak{H}_{D+1}^N\right)\rightarrow\infty,\\\nonumber
           \int_{Q_{D-1}}dV\left|\Xi_{D+1}^{N,V'}(V)\right|^2&=1.
     \end{align}
     We can also conclude that for bounded functions $f(V)$ on $Q_{D-1}$, we have
     \begin{equation}
     \lim_{ N\rightarrow\infty }\int_{Q_{D-1}}dV\left|\Xi_{D+1}^{N,V'}(V)\right|^2f(V)=f(V').
     \end{equation}
     Let us prove it as follows. Consider a region $\Delta$ around point $V'\in Q_{D-1}$ characterised by three infinitesimal angles $\Delta\theta_1, \Delta\theta_2, \Delta\gamma$, for which we have
      \begin{eqnarray}\label{outDelta}
      \left. \left|\Xi_{D+1}^{N,V'}(V)\right|^2\right|_{V\in Q_{D-1}\backslash\Delta}&\leq&\dim\left(\mathfrak{H}_{D+1}^N\right)\left(\frac{\cos\Delta\theta_1+e^{\mathbf{i}\Delta\gamma}\cos\Delta\theta_2}{2}\right)^N \overline{\left(\frac{\cos\Delta\theta_1+e^{\mathbf{i}\Delta\gamma}\cos\Delta\theta_2}{2}\right)}^N\\\nonumber & =&\dim\left(\mathfrak{H}_{D+1}^N\right)\left(\frac{\cos^2\Delta\theta_1+\cos^2\Delta\theta_2+2\cos\Delta\gamma\cos\Delta\theta_1\cos\Delta\theta_2}{4}\right)^N,
      \end{eqnarray}
and
\begin{eqnarray}\label{predelta}
\int_{Q_{D-1}}dV\left|\Xi_{D+1}^{N,V'}(V)\right|^2f(V)
=\int_{Q_{D-1}\backslash\Delta} dV\left|\Xi_{D+1}^{N,V'}(V)\right|^2f(V)+\int_{\Delta}dV\left|\Xi_{D+1}^{N,V'}(V)\right|^2f(V).
\end{eqnarray}

First, due to Eqs.\eqref{gamma}, \eqref{PV}, and \eqref{outDelta}, we have for $\Delta\theta_1,\Delta\theta_2\rightarrow0$ at large $N$
\begin{eqnarray}
\lim_{ N\rightarrow\infty}\int_{Q_{D-1}\backslash\Delta}dV\left|\Xi_{D+1}^{N,V'}(V)\right|^2f(V)&\leq&\max(|f(V)|)\int_{Q_{D-1}\backslash \Delta}dV\left|\Xi_{D+1}^{N,V'}(V)\right|^2\rightarrow0, \\\nonumber
\lim_{ N\rightarrow\infty}  \int_{\Delta} dV\left|\Xi_{D+1}^{N,V'}(V)\right|^2&\rightarrow&1,
\end{eqnarray}
where we used the fact that the righthand side of Eq.\eqref{outDelta} tends to zero in large $N$ limit, since the factor $(\frac{\cos^2\Delta\theta_1+\cos^2\Delta\theta_2+2\cos\Delta\gamma\cos\Delta\theta_1\cos\Delta \theta_2}{4})^N$ in Eq.\eqref{outDelta} decreases exponentially with $N\rightarrow \infty$, while another factor $\dim\left(\mathfrak{H}_{D+1}^N\right)$ in Eq.\eqref{outDelta} only increases polynomially in $N$.
Second, for arbitrary bounded functions $f(V)$ whose derivative is finite at every point of $Q_{D-1}$, we have
\begin{equation}
\lim_{\Delta\theta_1,\Delta\theta_2\rightarrow0}\int_{\Delta} dVf(V)\left|\Xi_{D+1}^{N,V'}(V)\right|^2\rightarrow f(V')\int_{\Delta} dV\left|\Xi_{D+1}^{N,V'}(V)\right|^2.
\end{equation}

Then, based on the above two points and \eqref{predelta}, we can immediately conclude that
\begin{eqnarray}\label{deltaVV}
\lim_{ N\rightarrow\infty}\int_{Q_{D-1}}dV\left|\Xi_{D+1}^{N,V'}(V)\right|^2f(V)=f(V'),
\end{eqnarray}
 which finishes our proof. In addition, an error estimation is given in the appendix A, which shows that the error of the above equation can be bounded by $\mathcal{E}\sim N^{-\frac{\beta}{2}}$ for a proper choice of $\Delta$ and $0<\beta<1$.
 A similar discussion can be given for
 \begin{eqnarray}
 \left|\Xi_{D+1}^{N,V'}(V)\right|&=&\sqrt{\dim(\mathfrak{H}_{D+1}^{N})}\left|\langle N,V'|N,V \rangle\right|\\\nonumber
 &=&\sqrt{\dim(\mathfrak{H}_{D+1}^{N})}\left(\frac{\cos^2\Delta\theta_1+\cos^2\Delta\theta_2 +2\cos\Delta\gamma\cos\Delta\theta_1\cos\Delta\theta_2}{4}\right)^{N/2},
 \end{eqnarray}
which means we also have
\begin{eqnarray}\label{deltaVV2}
\lim_{ N\rightarrow\infty}\int_{Q_{D-1}}dV\left|\Xi_{D+1}^{N,V'}(V)\right|f(V)=f(V').
\end{eqnarray}

This result can be extended to a more general case, i.e., the coherent intertwiner constructed by the $SO(D+1)$ coherent state. Let us consider the gauge fixed simple coherent intertwiners $|\vec{N}, \vec{V}\rangle:=\otimes_{\imath=1}^{n_v}|N_\imath, V_\imath\rangle$ which can be labelled to a $n_v$ valent vertex \cite{long2019coherent}. The inner product $\langle \vec{N}, \vec{V}|\vec{N}, \vec{V}'\rangle$ of two arbitrary gauge fixed simple coherent intertwiners can be given by
\begin{eqnarray}\label{NVNV'}
  \langle\vec{N},\vec{V}|\vec{N},\vec{V}'\rangle &=&\prod_{\imath=1}^{n_v}\left(\frac{\cos\theta^\imath _1+e^{\mathbf{i}\gamma^\imath}\cos\theta^\imath_2}{2}\right)^{N_\imath}\cdot e^{-\mathbf{i}N_{\imath}\phi_{\imath}}\\\nonumber
  &=&\left( \frac{\cos^2\theta_1^\imath+\cos^2\theta^\imath_2 +2\cos\gamma^\imath\cos\theta^\imath_1\cos\theta^\imath_2}{4}\right)^{N_\imath/2}e^{\mathbf{i}N_\imath\varphi_\imath} e^{\mathbf{i}N_{\imath}\phi_{\imath}},
\end{eqnarray}
wherein $(\theta_1^\imath, \theta^\imath_2, \gamma^\imath)$ is the set of angles between the bi-vectors $V_\imath$ and $V'_\imath$ (see the introduction below \eqref{gamma}) and $\varphi_\imath:=\arctan\left(\frac{\sin\gamma^\imath\cos\theta^\imath_2}{\cos\theta^\imath _1+\cos\gamma^\imath\cos\theta^\imath_2}\right)$. It is easy to see that the inner product $\langle\vec{N},\vec{V}|\vec{N},\vec{V}'\rangle$ has maximal value 1 at $\vec{V}=\vec{V}'$ and it decreases exponentially with $\vec{N}\rightarrow\vec{\infty}$ if $\vec{V}\neq\vec{V}'$. Then, similar to this discussion, we can give
\begin{eqnarray}\label{NVNV}
  \langle\vec{N},\vec{V}|g^{\otimes n_v}|\vec{N},\vec{V}\rangle
  &=&\prod_{\imath=1}^{n_v} \langle N_\imath,V_\imath|g|N_\imath,V_\imath\rangle\\\nonumber
  &=&\prod_{\imath=1}^{n_v}\left(\frac{\cos\theta^\imath _1(g)+e^{\mathbf{i}\gamma^\imath(g)}\cos\theta^\imath_2(g)}{2}\right)^{N_\imath}\cdot e^{\mathbf{i}N_{\imath}\phi_{\imath}(g)}\\\nonumber
  &=& \prod_{\imath=1}^{n_v}\chi_{N_\imath}^\imath(g)e^{\mathbf{i}N_\imath\varphi_\imath(g)}e^{\mathbf{i}N_{\imath}\phi_{\imath}(g)}
\end{eqnarray}
 wherein $(\theta_1^\imath(g), \theta^\imath_2(g), \gamma^\imath(g))$ is the set of angles between the bi-vectors $V_\imath$ and $gV_\imath g^{-1}$, $\varphi_\imath(g):=\arctan\left(\frac{\sin\gamma^\imath(g)\cos\theta^\imath_2(g)}{\cos\theta^\imath _1(g)+\cos\gamma^\imath(g)\cos\theta^\imath_2(g)}\right)$, and
 \begin{equation}
 \chi_{N_\imath}^\imath(g):=\chi_{N_\imath}^\imath(\theta^\imath _1(g),\theta^\imath _2(g),\gamma^\imath (g)):= \left( \frac{\cos^2\theta_1^\imath(g)+\cos^2\theta^\imath_2(g) +2\cos\gamma^\imath(g)\cos\theta^\imath_1(g)\cos\theta^\imath_2(g)}{4}\right)^{N_\imath/2}.
 \end{equation}
  Also, we can see that the function $\chi_{N_\imath}^\imath(g)$ is peaked at the subgroup $(SO(2)\times SO(D-1))_{V^\imath}$ which fixes the bi-vector $V^\imath$ and the peakedness becomes sharp in the large $N_\imath$ limit. Notice that the function $c_{\chi}^{\vec{N}}\prod_{\imath=1}^{n_v}\chi_{N_\imath}^\imath(g)$ satisfies
  \begin{eqnarray}
\lim_{\vec{N}\rightarrow\vec{\infty}}\left.c_{\chi}^{\vec{N}}\prod_{\imath=1}^{n_v}\chi_{N_\imath}^\imath(g)\right|_{g=\text{Id.}} &\rightarrow& \infty, \\\nonumber
     \int_{SO(D+1)}dg \, c_{\chi}^{\vec{N}}\prod_{\imath=1}^{n_v}\chi_{N_\imath}^\imath(g) &=&1,
  \end{eqnarray}
   with $c_{\chi}^{\vec{N}}=\frac{1}{\int_{SO(D+1)}dg\prod_{\imath=1}^{n_v}\chi_{N_\imath}^\imath(g)}$.
  Hence, following the same procedures as in the above proof, we can also show that for a bounded function $f(g)$ on SO$(D+1)$, we have
  \begin{equation}\label{deltag}
\lim_{ \vec{N}\rightarrow\vec{\infty}}\int_{SO(D+1)}dg \, c_{\chi}^{\vec{N}}\prod_{\imath=1}^{n_v}\chi_{N_\imath}^\imath(g)f(g) =\left.f(g)\right|_{g=\text{Id.}},
 \end{equation}
which implies that
  $c_{\chi}^{\vec{N}}\prod_{\imath=1}^{n_v}\chi_{N_\imath}^\imath(g)$ tends to a delta distribution on SO$(D+1)$ in the large $N_\imath$ limit. Finally, let us look at Eq.\eqref{NVNV} and notice that $e^{\mathbf{i}N_\imath\varphi_\imath(g)}e^{\mathbf{i}N_{\imath}\phi_{\imath}(g)}$ is a phase factor with frequency $N_\imath$. A similar result can be given for
  $\delta_\chi^{\vec{N},\vec{V}}(g):=\prod_{\imath=1}^{n_v}\chi_{N_\imath}^\imath(g)e^{\mathbf{i}N_\imath\varphi_\imath(g)}e^{\mathbf{i}N_{\imath}\phi_{\imath}(g)} $, that is
  \begin{equation}\label{deltae}
  \lim_{ \vec{N}\rightarrow\vec{\infty}}\int_{SO(D+1)}dg \, \delta_\chi^{\vec{N},\vec{V}}(g)f(g)=\left.f(g)\right|_{g=\text{Id.}}\cdot \lim_{ \vec{N}\rightarrow\vec{\infty}}\int_{SO(D+1)} dg \, \delta_\chi^{\vec{N},\vec{V}}(g).
  \end{equation}

     \item \textit{The coherent state representation is appropriate for describing operators. For an operator $\hat{O}$ which is a function of $\hat{J}^{ij}$, we can define its symbols $\mathbf{P}_{\hat{O}}(V)$ and $ \mathbf{Q}_{\hat{O}}(V)$ by}
         \begin{eqnarray}\label{Ps}
         \hat{O}&=&\int_{Q_{D-1}}d\mu_N(V)\mathbf{P}_{\hat{O}}(V)|N,V\rangle\langle N,V|,\quad ~~ d\mu_N(V):=\dim\left(\mathfrak{H}_{D+1}^N\right)dV,\\\nonumber
         \mathbf{Q}_{\hat{O}}(V)&=&\langle N,V|\hat{O}|N,V\rangle.
         \end{eqnarray}
         Properties of these symbols can be generalized from previous works \cite{GeneralizedCoherentStates} about coherent states of other Lie groups. The two symbols are consistent with each other in the large $N$ limit, i.e.
         \begin{eqnarray}
        \lim_{ N\rightarrow\infty}\mathbf{Q}_{\hat{O}}(V')&=&\lim_{ N\rightarrow\infty}\langle N,V'|\hat{O}|N,V'\rangle\\\nonumber
            &=& \lim_{ N\rightarrow\infty} \int_{Q_{D-1}}d\mu_N(V)\mathbf{P}_{\hat{O}}(V)|\langle N,V|N,V'\rangle|^2
            \\\nonumber
            &=& \lim_{ N\rightarrow\infty} \int_{Q_{D-1}}dV\mathbf{P}_{\hat{O}}(V)\left|\Xi_{D+1}^{N,V'}(V)\right|^2
            \\\nonumber
            &=&\mathbf{P}_{\hat{O}}(V'),
         \end{eqnarray}
         where we used \eqref{deltaVV}.
\end{enumerate}

 In LQG, the action of flux operators on quantum states is closely related to the action of $X^{IJ}$ on states in $\mathfrak{H}^N_{D+1}$, and it plays a key role in the study of spatial geometric operators. Due to their action as derivatives, it is worth to discuss the behaviour of the derivative of the matrix element functions on SO$(D+1)$ evaluated in Perelomov coherent states. Let us choose an orthogonal basis of the bi-vector space as $\{V^{IJ}, \{\bar{V}^{IJ}\}, \{V_{\perp}^{IJ}\}\}$, where  $\{\bar{V}^{IJ}\}$ is composed by the elements which commute with $V^{IJ}$, and $\{V_{\perp}^{IJ}\}$ represents the remaining elements. Now, we can show that,
   \begin{eqnarray}\label{NVXg}
   &&V_{IJ}\langle N,V|X^{IJ}g|N,V\rangle\\\nonumber
   &=&\mathbf{i}N\langle N,V|g|N,V\rangle=\mathbf{i}N\left(\frac{\cos\theta_1(g) +e^{\mathbf{i}\gamma(g)}\cos\theta_2(g)}{2}\right)^Ne^{\mathbf{i}N\phi(g)},\\
   \\\nonumber
  && \bar{V}_{IJ}\langle N,V|X^{IJ}g|N,V\rangle=0,\\
  \\\nonumber
   &&V^{\perp}_{IJ}\langle N,V|X^{IJ}g|N,V\rangle\\\nonumber
   &=&\frac{1}{2}\Theta_1(\theta_1(g),\theta_2(g),\gamma(g))N\sin\theta_1(g)\left(\frac{\cos\theta_1(g) +e^{\mathbf{i}\gamma(g)}\cos\theta_2(g)}{2}\right)^{(N-1)}e^{\mathbf{i}N\phi(g)}\\\nonumber
  &&+\frac{1}{2}\Theta_2(\theta_1(g),\theta_2(g),\gamma(g))Ne^{\mathbf{i}\gamma(g)} \sin\theta_2(g)\left(\frac{\cos\theta_1(g) +e^{\mathbf{i}\gamma(g)}\cos\theta_2(g)}{2}\right)^{(N-1)}e^{\mathbf{i}N\phi(g)}\\\nonumber
  &&+\frac{1}{2}\Theta_\gamma(\theta_1(g),\theta_2(g),\gamma(g))Ne^{\mathbf{i}\gamma(g)}\cos\theta_2(g) \left(\frac{\cos\theta_1(g) +e^{\mathbf{i}\gamma(g)}\cos\theta_2(g)}{2}\right)^{(N-1)}e^{\mathbf{i}N\phi(g)}\\\nonumber
  &=:&N\Psi_1\left(\theta_1(g),\theta_2(g),\gamma(g)\right)\left(\frac{\cos\theta_1(g) +e^{\mathbf{i}\gamma(g)}\cos\theta_2(g)}{2}\right)^{(N-1)}e^{\mathbf{i}N\phi(g)},
  \end{eqnarray}
  where
  \begin{eqnarray}
   \Theta_1(\theta_1(g),\theta_2(g),\gamma(g)) &:=&V^{\perp}_{IJ}\theta_1(X^{IJ}g)=V^{\perp}_{IJ}\frac{d}{dt}\theta_1(\exp(tX^{IJ})g), \\\nonumber
   \Theta_2(\theta_1(g),\theta_2(g),\gamma(g))&:=& V^{\perp}_{IJ}\theta_2(X^{IJ}g)=V^{\perp}_{IJ}\frac{d}{dt}\theta_2(\exp(tX^{IJ})g),
  \end{eqnarray}
and
\begin{eqnarray}\label{Thetagamma}
	\Theta_\gamma(\theta_1(g),\theta_2(g),\gamma(g)):=V^{\perp}_{IJ}\gamma(X^{IJ}g) =V^{\perp}_{IJ}\frac{d}{dt}\gamma(\exp(tX^{IJ})g),
\end{eqnarray}
which satisfies
  \begin{equation}
  \left.\Theta_\gamma(\theta_1(g),\theta_2(g),\gamma(g))\right|_{\theta_1=\theta_2=0}=0
  \end{equation}
  based on \eqref{gamma}. Let us define $f'_{N,V}(g):=\frac{1}{N}V_{IJ}\langle N,V|X^{IJ}g|N,V\rangle$, $f'_{N,\bar{V}}(g):=\frac{1}{N}\bar{V}_{IJ}\langle N,V|X^{IJ}g|N,V\rangle$ and $f'_{N,V_{\perp}}(g):=\frac{1}{N}\bar{V}^{\perp}_{IJ}\langle N,V|X^{IJ}g|N,V\rangle$. We conclude that
  \begin{enumerate}
  	\item $f'_{N,V}(g)$ is sharply peaked at $\theta_1(g)=\theta_2(g)=0$ for large $N$ and $\left.f'_{N,V}(g)\right|_{\theta_1(g)=\theta_2(g)=0}=\mathbf{i}e^{\mathbf{i}N\phi(g)}$.
	\item $f'_{N,\bar{V}}(g)=0$.
	\item $\lim_{N\rightarrow\infty}f'_{N,V_{\perp}}(g)\rightarrow0$, which follows from the fact that $\left(\frac{\cos\theta_1(g) +e^{\mathbf{i}\gamma(g)}\cos\theta_2(g)}{2}\right)^{(N-1)}$ in $f'_{N,V_{\perp}}(g)$ is sharply peaked at $\theta_1(g)=\theta_2(g)=0$, while $\sin\theta_1(g)$, $\sin\theta_2(g)$ and $\Theta_\gamma(\theta_1(g),\theta_2(g),\gamma(g))$ vanish at $\theta_1(g)=\theta_2(g)=0$, and also their derivatives are finite near $\theta_1(g)=\theta_2(g)=0$.
	\end{enumerate}
	 Similar discussion and results can be given for $f''_{N,V_1,V_2}(g):=\frac{1}{N^2}V_{1KL}V_{2IJ}\langle N,V|X^{KL}X^{IJ}g|N,V\rangle$ with $V_{1}^{KL},V_{2}^{IJ}\in\{V^{IJ}, \{\bar{V}^{IJ}\}, \{V_{\perp}^{IJ}\}\}$ and higher order derivatives
 \begin{equation}
 f^{[n]}_{N,\{V_1,...,V_n\}}(g):=\frac{1}{N^n}V_{1IJ}V_{2I'J'}...V_{nKL}\langle N,V|X^{IJ}X^{I'J'}...X^{KL}g|N,V\rangle
 \end{equation}
 with $V_{1IJ},V_{2I'J'},...,V_{nKL}\in\{V^{IJ}, \{\bar{V}^{IJ}\}, \{V_{\perp}^{IJ}\}\}$ and $n$ being a finite positive integer satisfying $n\ll N$. Let us consider three kinds of choices of $\{V_1,...,V_n\}$, they are (i) $V_{1IJ}=V_{2IJ}=...=V_{nIJ}=V^{IJ}$; (ii) $V_{\ell IJ}\in \{\bar{V}^{IJ}\}, 1\leq\ell\leq n$ and $V_{1 IJ}=V_{2IJ}=...=V_{(\ell-1)IJ}=V^{IJ}$ ; (iii) The other choices of $\{V_1,...,V_n\}$. We discuss these three choices separately.
 \begin{enumerate}
   \item For the choice (i), we have
 \begin{equation}
 f^{[n]}_{N,\{V_1,...,V_n\}}(g)=(\mathbf{i})^n\langle N,V|g|N,V\rangle,
 \end{equation}
 which is sharply peaked at $\theta_1(g)=\theta_2(g)=0$ for large $N$ and $\left.f^{[n]}_{N,\{V_1,...,V_n\}}(g)\right|_{\theta_1(g)=\theta_2(g)=0} =(\mathbf{i})^ne^{\mathbf{i}N\phi(g)}$.
   \item For the choice (ii), we have
    \begin{equation}
 f^{[n]}_{N,\{V_1,...,V_n\}}(g)=0.
 \end{equation}
   \item For the choice (iii), the properties of $f^{[n]}_{N,\{V_1,...,V_n\}}(g)$ can be analyzed as follows. Firstly, the value of $f^{[n]}_{N,\{V_1,...,V_n\}}(g)$ at ${\theta_1(g)=\theta_2(g)=0}$ is given by
     \begin{equation}
 \left.f^{[n]}_{N,\{V_1,...,V_n\}}(g)\right|_{\theta_1(g)=\theta_2(g)=0}=\frac{1}{N^n}V_{1IJ}V_{2I'J'}...V_{nKL}\langle N,V|X^{IJ}X^{I'J'}...X^{KL}|N,V\rangle.
 \end{equation}
 Notice that $V_{1IJ}V_{2I'J'}...V_{nKL}\langle N,V|X^{IJ}X^{I'J'}...X^{KL}|N,V\rangle$ takes the value $0$ or is a polynomial in $N$ with degree less than $n$ for choice (iii) of $\{V_1,...,V_n\}$, so that one has
 \begin{equation}\label{fnNV0} \left.f^{[n]}_{N,\{V_1,...,V_n\}}(g)\right|_{\theta_1(g)=\theta_2(g)=0}=0\,\, \text{or}\,\sim\frac{1}{N^{\iota}}
 \end{equation}
with $1\leq\iota< n$.
 Secondly, based on Eqs.\eqref{NVXg}-\eqref{Thetagamma} and the fact that $n$ is a finite positive integer, we know that $f^{[n]}_{N,\{V_1,...,V_n\}}(g)$ must be a sum of finite terms as
  \begin{equation}\label{fnNViii}
f^{[n]}_{N,\{V_1,...,V_n\}}(g)=\sum_{\tilde{N}} \mathfrak{F}_{\tilde{N}}(\theta_1(g),\theta_2(g),\gamma(g)) \left(\frac{\cos\theta_1(g) +e^{\mathbf{i}\gamma(g)}\cos\theta_2(g)}{2}\right)^{\tilde{N}}e^{\mathbf{i}N\phi(g)},
 \end{equation}
 with $N-n<\tilde{N}<N$ and $\mathfrak{F}_{\tilde{N}}(\theta_1(g),\theta_2(g),\gamma(g))$ being a bounded function whose derivative is finite near ${\theta_1(g)=\theta_2(g)=0}$. Now it is easy to see $\left.\sum_{\tilde{N}}\mathfrak{F}_{\tilde{N}}(\theta_1(g),\theta_2(g),\gamma(g)) \right|_{\theta_1(g)=\theta_2(g)=0}=0\,\,\text{or}\,\sim\frac{1}{N^{\iota}}$ based on Eqs.\eqref{fnNV0} and \eqref{fnNViii}. Then, notice that the factor $\left(\frac{\cos\theta_1(g) +e^{\mathbf{i}\gamma(g)}\cos\theta_2(g)}{2}\right)^{\tilde{N}}$ in Eq.\eqref{fnNViii} is sharply peaked at ${\theta_1(g)=\theta_2(g)=0}$ if $N$ is large and $n\ll N$, so that we can immediately conclude that
 \begin{equation}
\lim_{N\rightarrow\infty, n \ll N}f^{[n]}_{N,\{V_1,...,V_n\}}(g)=0
 \end{equation}
 for the choice (iii) of $\{V_1,...,V_n\}$.
 \end{enumerate}
  Now, based on the above discussion, we can conclude the last property of $SO(D+1)$ Perelomov coherent states in this section as a theorem, which reads,
  \begin{itemize}
    \item \textbf{Theorem.} \textit{The function $f^{[n]IJI'J'...KL}_{N}(g):=\frac{1}{N^n}\langle N,V|\overbrace{X^{IJ}X^{I'J'}...X^{KL}}^{\text{with}\ n-\text{tuple}\ X^{IJ}}g|N,V\rangle$ is a tensor valued function on $SO(D+1)$, which is sharply peaked at the maximum subgroup $\left(SO(2)\times SO(D-1)\right)_V$ that fixes the bi-vector $V$ in the limit $n \ll N, N\rightarrow \infty$, that is,}
         \begin{equation}\label{lastproperty1}
f^{[n]IJI'J'...KL}_{N}(g)=(2\mathbf{i})^nV^{IJ}V^{I'J'}...V^{KL}\langle N,V|g|N,V\rangle+\mathcal{O}^{[n]IJI'J'...KL}_{N}(g)
 \end{equation}
 \textit{with all of the components of $\mathcal{O}_{N}^{[n]IJI'J'...KL}(g)$ tending to zero in the limit $n\ll N$ and $N\rightarrow \infty$. By denoting $V'=g'Vg'^{-1}$ and defining}
 \begin{equation}
 f^{[n]IJI'J'...KL}_{N,V ,V'}(g):=\frac{1}{N^n}\langle N,V|\overbrace{X^{IJ}X^{I'J'}...X^{KL}}^{\text{with}\ n-\text{tuple}\ X^{IJ}}g|N,V'\rangle,
 \end{equation}
  \textit{ the above statement can be extended to more general case as}

    \begin{equation}\label{lastproperty2}
f^{[n]IJI'J'...KL}_{N,V ,V'}(g)=(2\mathbf{i})^nV^{IJ}V^{I'J'}...V^{KL}\langle N,V|g|N,V'\rangle+\mathcal{O}^{[n]IJI'J'...KL}_{N,V ,V'}(g)
 \end{equation}
\textit{with all of the components of $\mathcal{O}_{N,V ,V'}^{[n]IJI'J'...KL}(g)$ tending to zero in the limit $n\ll N$ and $N\rightarrow \infty$.}
 \end{itemize}
 This theorem will be very useful in the calculation of expectation values of geometric operators, which will be illustrated in the next section.

\section{Perelomov coherent states of SO$(D+1)$ in all dimensional loop quantum gravity}\label{sec:GeomOperators}
\subsection{Simple coherent intertwiner}
The Perelomov coherent states of SO$(D+1)$ are indispensible in the construction of simple coherent intertwiners in all dimensional loop quantum gravity, which are used to weakly solve the anomalous quantum vertex simplicity constraints \cite{long2019coherent}. The resulting spin network states, equipped with gauge invariant (or gauge fixed) simple coherent intertwiners, are constructed by labelling each edge of a closed graph with a simple representation of SO$(D+1)$ and each vertex with a simple coherent intertwiner \cite{Bodendorfer:Qu, long2019coherent}. More precisely, such weakly simple spin network states are linear combinations of products of matrix element functions on several copies of SO$(D+1)$. The matrix element functions are selected by Perelomov coherent states in the simple representation space of SO$(D+1)$, which take the form $\Xi_{D+1}^{N,V,V'}(g):=\sqrt{\dim(\mathfrak{H}^N_{D+1})}\langle N,V|g|N,V'\rangle$. Thus, it is worth to discuss the properties of these special functions. In LQG, the flux operators act on the related matrix element functions as right (or left) invariant vector fields on SO$(D+1)$ as
\begin{equation}\label{flux1}
\hat{F}^{IJ}\circ \Xi_{D+1}^{N,V,V'}(g)=\mathbf{i}\hbar\beta\kappa R^{IJ}\circ \Xi_{D+1}^{N,V,V'}(g) =\frac{1}{2}\mathbf{i}\hbar\beta\kappa \Xi_{D+1}^{N,V,V'}\left(X^{IJ}g \right)
\end{equation}
where $R^{IJ}$ is the right invariant vector field on SO$(D+1)$ which is defined by its action on a function $f(g)$ on SO$(D+1)$ as $R^{IJ}\circ f(g):=\frac{d}{dt}f(e^{\frac{1}{2}tX^{IJ}}g)|_{t=0}$.
The expectation value of $\hat{F}^{IJ}$ for this function is given by
\begin{eqnarray}\label{flux2}
\langle N,V,V'|\hat{F}^{IJ}| N,V,V'\rangle&:=&\int_{SO(D+1)}dg\overline{\Xi_{D+1}^{N,V,V'}(g)}\hat{F}^{IJ}\circ \Xi_{D+1}^{N,V,V'}(g)\\\nonumber
&=&\frac{1}{2}\mathbf{i}\hbar\beta\kappa \int_{SO(D+1)}dg\overline{\Xi_{D+1}^{N,V,V'}(g)} \Xi_{D+1}^{N,V,V'}(X^{IJ}g)\\\nonumber
&=&\frac{1}{2}\mathbf{i}\hbar\beta\kappa \langle N,V|X^{IJ}|N,V\rangle,
\end{eqnarray}
where we used the fact that \cite{vilenkin2013representation}
\begin{equation}\label{NgNgN}
\int_{SO(D+1)}dg\overline{\langle N,\mathbf{M}_1|g|N,\mathbf{M}_2\rangle}\langle N,\mathbf{M}'_1|g|N,\mathbf{M}'_2\rangle=\frac{1}{\dim(\mathfrak{H}^N_{D+1})} \delta_{\mathbf{M}_1,\mathbf{M}'_1}\delta_{\mathbf{M}_2,\mathbf{M}'_2}.
\end{equation}

Based on this property, we can further focus on the simple coherent intertwiner which involves $n_v$ edges linked to a vertex \cite{long2019coherent}.  Notice that the simple coherent intertwiner space is a subspace of the direct product $\otimes_{\imath=1}^{n_v}\mathfrak{H}_{D+1}^{N_\imath}$, and simple coherent intertwiners can be written as
\begin{equation}
|\vec{N},\vec{V}\rangle:=\otimes_{\imath=1}^{n_v}|N_{\imath},V_\imath\rangle
\end{equation}
in the gauge fixed case, and as
\begin{equation}
||\vec{N},\vec{V}\rangle:=\int_{SO(D+1)}dg\otimes_{\imath=1}^{n_v}g|N_{\imath},V_\imath\rangle
\end{equation}
in the gauge invariant case, wherein the labelling bi-vectors $V^{IJ}_{\imath}$ satisfy the classical simplicity constraint $V^{[IJ}_{\imath}V^{KL]}_{\jmath}=0$ and the closure condition  $\sum_{\imath=1}^{n_v}N_\imath V_\imath^{IJ}=0$ \cite{Long:2020agv}. The simple coherent intertwiners weakly solve the quantum vertex simplicity constraints as follows. Consider the tensor valued operator $X_{\jmath_1}^{IJ}X_{\jmath_2}^{KL}$ whose totally asymmetry part $X_{\jmath_1}^{[IJ}X_{\jmath_2}^{KL]}$ is the quantum vertex simplicity constraints operator, and a geometric operator $\hat{G}(...,X_{\jmath_1}^{IJ}X_{\jmath_2}^{KL},...)$ which contain the factor $X_{\jmath_1}^{IJ}X_{\jmath_2}^{KL}$. A state weakly solve the quantum vertex simplicity constraints means that the expectation value of  $X_{\jmath_1}^{[IJ}X_{\jmath_2}^{KL]}$ in this state is infinite small relative to the contribution of the factor $X_{\jmath_1}^{IJ}X_{\jmath_2}^{KL}$ to the expectation value of $\hat{G}(...,X_{\jmath_1}^{IJ}X_{\jmath_2}^{KL},...)$ in this state. Usually, this contribution has the tensor norm $N_{\jmath_1}N_{\jmath_2}$ for the state $||\vec{N},\vec{V}\rangle$ (see the volume operator as an example in next subsection). Then, it is easy to check that the gauge fixed simple coherent intertwiners provide a weak solution space to the quantum vertex simplicity constraints as \cite{long2019coherent}
\begin{equation}
\langle \vec{N},\vec{V}|X_{\jmath_1}^{[IJ}X_{\jmath_2}^{KL]}|\vec{N},\vec{V}\rangle=0.
\end{equation}
Also, it has been shown that  \cite{long2019coherent}
\begin{equation}
\lim_{N\rightarrow\infty}\frac{\langle \vec{N},\vec{V}||X_{\jmath_1}^{[IJ}X_{\jmath_2}^{KL]}||\vec{N},\vec{V}\rangle} {N_{\jmath_1}N_{\jmath_2}\langle \vec{N},\vec{V}||\vec{N},\vec{V}\rangle}=0,
\end{equation}
which means the gauge invariant simple coherent intertwiners provide a weak solution space to the quantum vertex simplicity constraints in large $N$ limit at least.
We can also check that the non-diagonal elements of the quantum vertex simplicity constraint operator vanish weakly as
\begin{equation}
\lim_{N\rightarrow\infty}\frac{\langle \vec{N},\vec{V}|X_{\jmath_1}^{[IJ}X_{\jmath_2}^{KL]}|\vec{N},\vec{V'}\rangle}{N_{\jmath_1}N_{\jmath_2}} =\lim_{N\rightarrow\infty}(-2\mathbf{i})^2V^{[IJ}_{\jmath_1}V^{KL]}_{\jmath_2}\langle \vec{N},\vec{V}|\vec{N},\vec{V'}\rangle=0
\end{equation}
and
\begin{equation}
\lim_{N\rightarrow\infty}\frac{\langle \vec{N},\vec{V}||X_{\jmath_1}^{[IJ}X_{\jmath_2}^{KL]}||\vec{N},\vec{V'}\rangle}{N_{\jmath_1}N_{\jmath_2} \sqrt{\langle \vec{N},\vec{V}||\vec{N},\vec{V}\rangle\langle \vec{N},\vec{V'}||\vec{N},\vec{V'}\rangle}}=0
\end{equation}
by using the theorem in section 3.
 Such formulations of simple coherent intertwiners make sure that the properties of a single SO$(D+1)$ coherent state can be generalized to the case of all dimensional LQG.
\subsection{Geometric operators}
  Spin network states labelled with gauge fixed simple coherent intertwiners are good coherent states for flux operators due to being products of matrix element functions $\Xi_{D+1}^{N,V,V'}(g)$. Also, spin network states labelled with gauge invariant simple coherent intertwiners can be regarded as good coherent states for the gauge invariant spatial geometric operators which can in several cases be build using only flux operators \cite{Bodendorfer:Qu, long2020operators}. In detail, these general spatial geometric operators are constructed by writing classical geometric quantities with classical fluxes and then replacing them with flux operators \cite{Bodendorfer:Qu, long2020operators}.
In this process, it is often necessary to compute a root of a finite polynomial of flux operators, which is done by an appeal to the spectral theorem. In computations of expectation values, we circumvent this step by arguing that in the large $N$ limit, we can exchange taking the root and computing the expectation value.
We will give the details of such a calculation for the $D$-volume operator (with $D$ odd) as an example in the following.

Let us first give the explicit expression of the $D$-volume operator (with $D$ odd). Consider a sub-Hilbert space $\mathcal{H}_\gamma$ which are composed of the cylindrical functions constructed on a graph $\gamma$, then the $D$-volume operator (with $D$ odd) for an infinitely small region $\square_\epsilon$ of coordinate size $\sim \epsilon^D$ in $\mathcal{H}_\gamma$ is given by \cite{Bodendorfer:Qu}
\begin{eqnarray}
\hat{V}_{\square_\epsilon}  &=&\int_{\square_\epsilon} d^Dp\hat{V}(p)_\gamma,\\\nonumber
  \hat{V}(p)_\gamma &=& (\hbar\kappa\beta)^{\frac{D}{D-1}}\sum_{v\in V(\gamma)}\delta^D(p,v)\hat{V}_{v,\gamma}, \\\nonumber
  \hat{V}_{v,\gamma} &=& |\frac{\mathbf{i}^D}{D!}\sum_{e_1,...,e_D\in E(\gamma),e_1\cap...\cap e_D=v}s(e_1,...,e_D)\hat{q}_{e_1,...,e_D}|^{\frac{1}{D-1}}, \\\nonumber
   \hat{q}_{e_1,...,e_D}&=&  \frac{1}{2}\epsilon_{IJI_1J_1I_2J_2...I_nJ_n}R_e^{IJ}R_{e_1}^{I_1K_1}R_{e'_1K_1}^{J_1}... R_{e_n}^{I_nK_n}R_{e'_nK_n}^{J_n},
   \end{eqnarray}
   where we re-labelled the edges $\{e_1,...,e_D\}$  as $\{e,e_1,e'_1,...,e_n,e'_n\}$ in the last line,  $\epsilon_{IJI_1J_1I_2J_2...I_nJ_n}$ is the Levi-Civita symbol in the internal space, and $R_e^{IJ}:=\frac{1}{2}\textrm{tr}((X^{IJ}h_e(A))^T\frac{\partial}{\partial h_e(A)})$ is the right invariant vector fields on $SO(D+1)\ni h_e(A)$ with $T$ representing transposition.
Let us denote
\begin{equation}\label{Qv}
\hat{Q}_{v,\gamma}:=\frac{\mathbf{i}^D}{D!}\sum_{e_1,...,e_D\in E(\gamma),e_1\cap...\cap e_D=v}s(e_1,...,e_D)\hat{q}_{e_1,...,e_D},
\end{equation}
so that $\hat{V}_{v,\gamma}=(\hat{Q}_{v,\gamma}^2)^{\frac{1}{2D-2}}$, where we should note the fact that $\mathbf{i}R_e^{IJ}$ is a real operator. In the following part of this section,  to proceed the calculation of expectation value of $(\hat{Q}_{v,\gamma}^2)^{\frac{1}{2D-2}}$, we will firstly deal with two obstacles, that are, (i) The $(2D-2)_{\text{th}}$ root over the operator $\hat{Q}_{v,\gamma}^2$; (ii) The action of operator $\hat{Q}_{v,\gamma}^2$ does not preserve the simple coherent intertwiner space. Then, with this two obstacle being overcome by introducing a projection operator and a Lemma, the calculation is transferred to calculating the expectation value of a polynomial of flux operator. Hence the further calculation can be done by using the property of Perelomov coherent states of SO$(D+1)$ which is concluded as the Theorem in section 3. Now, let us turning to the details of these discussion and calculations.

 In principle, to deal with $(2D-2)_{\text{th}}$ root over the operator $\hat{Q}_{v,\gamma}^2$, we need to find the eigenstates of the operator $\hat{Q}_{v,\gamma}^2$ and give its eigen-spectrum $\text{Spec}(\hat{Q}_{v,\gamma}^2)$, then the eigen-spectrum of $\hat{V}_{v,\gamma}=(\hat{Q}_{v,\gamma}^2)^{\frac{1}{2D-2}}$ will be given by $\text{Spec}(\hat{V}_{v,\gamma})=(\text{Spec}(\hat{Q}_{v,\gamma}^2))^{\frac{1}{2D-2}}$ for corresponding eigenstates. Unfortunately, it seems that the eigenstates of $(\hat{Q}_{v,\gamma}^2)$ are not lying in the simple coherent intertwiner space, because $(\hat{Q}_{v,\gamma}^2)$ is not commuting with the quantum vertex simplicity constraints. This would imply that eigenstates of the volume operator have no correct physical meaning. A possible way to solve this problem is to insert an orthogonal projection operator $\mathbb{P}_v^{\text{s}}$ into the solution space of the vertex simplicity constraints on both sides of $\hat{Q}_{v,\gamma}^2$ \cite{Bodendorfer:2011onthe}. Based on such a treatment, we will calculate the expectation value of $\hat{V}_{v,\gamma}=(\mathbb{P}_v^{\text{s}}\hat{Q}_{v,\gamma}^2\mathbb{P}_v^{\text{s}})^{\frac{1}{2D-2}}$ for the states labelled with simple coherent intertwiners. Suppose $v\in \gamma$ is a $n_v$-valent vertex and we denote by $|\gamma, (\vec{N},\vec{V})_v,...\rangle$ and $|\gamma, [\vec{N},\vec{V}]_v,...\rangle$ the spin-network states constructed on $\gamma$ with $v$ being labelled with the gauge invariant simple coherent intertwiner $||\vec{N},\vec{V}\rangle$ and the gauge fixed simple coherent intertwiner $|\vec{N},\vec{V}\rangle$ respectively. Then the expectation value of $\hat{V}_{v,\gamma}$ is defined by
\begin{equation}\label{epV}
\langle\hat{V}_{v,\gamma}\rangle
:=\frac{\langle\gamma, (\vec{N},\vec{V})_v,...|(\mathbb{P}_v^{\text{s}}\hat{Q}_{v,\gamma}^2\mathbb{P}_v^{\text{s}})^{\frac{1}{2D-2}}|\gamma, (\vec{N},\vec{V})_v,...\rangle} {\langle\gamma, (\vec{N},\vec{V})_v,...|\gamma, (\vec{N},\vec{V})_v,...\rangle}=\frac{\langle\gamma, (\vec{N},\vec{V})_v,...|(\hat{Q}_{v,\gamma}^2)^{\frac{1}{2D-2}}|\gamma, (\vec{N},\vec{V})_v,...\rangle} {\langle\gamma, (\vec{N},\vec{V})_v,...|\gamma, (\vec{N},\vec{V})_v,...\rangle},
\end{equation}
where we used the fact that $\hat{Q}_{v,\gamma}^2$ and $\mathbb{P}_v^{\text{s}}$ are both self-adjoint operators so that $\mathbb{P}_v^{\text{s}}\hat{Q}_{v,\gamma}^2\mathbb{P}_v^{\text{s}}$ and $\hat{Q}_{v,\gamma}^2$ are diagonalized in the simple coherent intertwiner space identically.
 The first obstacle to do this calculation is the $(2D-2)_{\text{th}}$ root of the operator $\hat{Q}_{v,\gamma}^2$. To overcome this problem, let us introduce the following Lemma which have been proven in Ref.\cite{Giesel_2007}.\\
 \
 \\
\textbf{Lemma.} (i) $1+qt-(1-q)t^2\leq(1+t)^q\leq1+qt$ for all $t\geq-1$ all $0<q<1$. (ii) Let $B_-\leq B\leq B_+$ be self-adjoint operators and set $\bar{B}:=\frac{(B_++B_-)}{2}$, $\delta B:=\frac{(B_+-B_-)}{4}$. Then for any states $\phi_1$ and $\phi_2$ in the common domain of all three operators we have
\begin{equation}
|\text{Re}\langle \phi_1|(B-\bar{B})|\phi_2\rangle|, |\text{Im}\langle \phi_1|(B-\bar{B})|\phi_2\rangle|\leq \langle \phi_1|\Delta B|\phi_1\rangle+\langle \phi_2|\Delta B|\phi_2\rangle.
\end{equation}

To apply this lemma, let us set $B=(\hat{Q}_{v,\gamma}^2)^q$ with $\hat{Q}_{v,\gamma}^2$ being positive semi-definite and self-adjoint operator. We have $B=\lambda^q(1+\frac{\hat{Q}_{v,\gamma}^2-\lambda}{\lambda})^q$ for arbitrary $\lambda>0$, and it is obviously that $z:=\frac{\hat{Q}_{v,\gamma}^2-\lambda}{\lambda}\geq-1$. Then we can set $B_+=\lambda^q(1+qz)$, $B_-=\lambda^q(1+qz-(1-q)z^2)$. Given a coherent state $\phi_1$ let us set $\lambda=\langle\phi_1|\hat{Q}_{v,\gamma}^2|\phi_1\rangle$ and we have $\langle\phi_1|B_+|\phi_1\rangle=\lambda^q$, $\langle\phi_1|B_-|\phi_1\rangle=\lambda^q(1-(1-q)(\frac{\langle\phi_1|\hat{Q} _{v,\gamma}^4|\phi_1\rangle}{\lambda^2}-1)^2)$. By using the above lemma we can immediately conclude that Eq.\eqref{epV} can be given by $\langle\hat{Q}_{v,\gamma}^2\rangle^{\frac{1}{2D-2}}$ up to a correction which will vanish if $\langle\hat{Q}_{v,\gamma}^2\rangle^2-\langle\hat{Q}_{v,\gamma}^4\rangle=0$, where we define
\begin{equation}
\langle\hat{Q}_{v,\gamma}^2\rangle
:=\frac{\langle\gamma, (\vec{N},\vec{V})_v,...|\hat{Q}_{v,\gamma}^2|\gamma, (\vec{N},\vec{V})_v,...\rangle} {\langle\gamma, (\vec{N},\vec{V})_v,...|\gamma, (\vec{N},\vec{V})_v,...\rangle},\  \langle\hat{Q}_{v,\gamma}^4\rangle
:=\frac{\langle\gamma, (\vec{N},\vec{V})_v,...|\hat{Q}_{v,\gamma}^4|\gamma, (\vec{N},\vec{V})_v,...\rangle} {\langle\gamma, (\vec{N},\vec{V})_v,...|\gamma, (\vec{N},\vec{V})_v,...\rangle}.
\end{equation}
Now let us calculate the above two equations respectively. By using Eqs. \eqref{flux1}, \eqref{flux2} and the theorem in section 3, we can get
\begin{eqnarray}\label{epQ}
\langle\hat{Q}_{v,\gamma}^2\rangle
&\stackrel{\vec N\text{ large}}{=}&\frac{\sum_{\{e\}}\sum_{\{e\}'}s(\{e\})s(\{e\}') (\prod_{e_\imath\in\{e\}} N_{e_\imath})(\prod_{e_\jmath\in\{e\}'}N_{e_\jmath})\epsilon(\{e\})\epsilon(\{e\}')} {4(D!)^2}\\\nonumber
&=&  \left(\frac{\langle\gamma, [\vec{N},\vec{V}]_v,...|\hat{Q}_{v,\gamma}|\gamma, [\vec{N},\vec{V}]_v,...\rangle} {\langle\gamma, [\vec{N},\vec{V}]_v,...|\gamma, [\vec{N},\vec{V}]_v,...\rangle}\right)^2=([\hat{Q}_{v,\gamma}])^2
\end{eqnarray}
with $[\hat{Q}_{v,\gamma}]:=\frac{\langle\gamma, [\vec{N},\vec{V}]_v,...|\hat{Q}_{v,\gamma}|\gamma, [\vec{N},\vec{V}]_v,...\rangle} {\langle\gamma, [\vec{N},\vec{V}]_v,...|\gamma, [\vec{N},\vec{V}]_v,...\rangle}$,
 see details of the calculation in Appendix B, where $\{e\}$ and $\{e\}'$ are two choices of the set $\{e_1,...,e_D\}$ satisfying $e_1,...,e_D\in E(\gamma),e_1\cap...\cap e_D=v$, $N_e$ with $e\in \{e\}$ or $\{e\}'$ is the quantum number labelled to the edge $e$ which is determined by the intertwiner $||\vec{N},\vec{V}\rangle$ labelled to $v$, $\epsilon(\{e\}):=\epsilon_{I J...KL}V_{e_1}^{IJ}...V_{e_D}^{KL}|_{e_1,...,e_D\in\{e\}}$ and $\epsilon(\{e\}'):=\epsilon_{I J...KL}V_{e_1}^{I'J'}...V_{e_D}^{K'L'}|_{e_1,...,e_D\in\{e\}'}$ with $V_e, e\in \{e\}$ or $\{e\}'$ is the bi-vector labelled to the edge $e$ which is determined by the intertwiner $||\vec{N},\vec{V}\rangle$ labelled to $v$.
Similarly discussion can be given for $\langle\hat{Q}_{v,\gamma}^4\rangle$ and we find
\begin{eqnarray}\label{epQ4}
\langle\hat{Q}_{v,\gamma}^2\mathbb{P}_v^{\text{s}}\hat{Q}_{v,\gamma}^2\rangle
\stackrel{\vec N\text{ large}}{=}\langle\hat{Q}_{v,\gamma}^4\rangle
&\stackrel{\vec N\text{ large}}{=}&
[\hat{Q}_{v,\gamma}]^4.
\end{eqnarray}
 Now, with the application of the above lemma and Eqs.\eqref{epQ} and \eqref{epQ4}, we can immediately get
\begin{equation}
\langle\hat{V}_{v,\gamma}\rangle
\stackrel{\vec N\text{ large}}{=}
\langle\hat{Q}^2_{v,\gamma}\rangle^\frac{1}{2D-2}
\stackrel{\vec N\text{ large}}{=}([\hat{Q}_{v,\gamma}])^\frac{1}{D-1}.
\end{equation}

The above calculation can be extended to more general case. For a finite degree polynomial $\mathcal{P}(\hat{F})$ of the flux operator $\hat{F}_{v,\gamma}^{IJ}$ which acts on the vertex $v\in\gamma$. If $\mathcal{P}(\hat{F})$ is positive semi-definite, self-adjoint and invariant under the gauge transformation induced by Gaussian constraint, then the theorem in section 3 and above Lemma are applicable so that we can get
\begin{equation}
\langle(\mathcal{P}(\hat{F}))^q\rangle
\stackrel{\vec N\text{ large}}{=}
([\mathcal{P}(\hat{F})])^q
\end{equation}
with $0<q<1$, $\langle(\mathcal{P}(\hat{F}))^q\rangle:=\frac{\langle\gamma, (\vec{N},\vec{V})_v,...|(\mathcal{P}(\hat{F}))^q|\gamma, (\vec{N},\vec{V})_v,...\rangle} {\langle\gamma, (\vec{N},\vec{V})_v,...|\gamma, (\vec{N},\vec{V})_v,...\rangle}$ and $[\mathcal{P}(\hat{F})]:=\frac{\langle\gamma, [\vec{N},\vec{V}]_v,...|\mathcal{P}(\hat{F})|\gamma, [\vec{N},\vec{V}]_v,...\rangle} {\langle\gamma, [\vec{N},\vec{V}]_v,...|\gamma, [\vec{N},\vec{V}]_v,...\rangle}$.

\section{Conclusion and discussion}
To better explore the kinematic structure of all dimensional LQG, we studied the general properties of Perelomov type coherent states of SO$(D+1)$, which are building blocks of the so-called simple coherent intertwiners which weakly solve the anomalous quantum vertices simplicity constraint in large quantum number $N$ limit \cite{long2019coherent}. Based on these properties, we also calculated the expectation value of the D-volume operator in all dimensional LQG and made a certain extension of the result.

 For pedagogical purposes, we first discussed a particle moving on a $D$-sphere, whose quantum angular momentum algebra served as a more familiar perspective to realize the quantum flux algebra, and the corresponding representations satisfies the simplicity constraint acting on edges. In these representations, the flux operators act on harmonic homogeneous functions on the $D$-sphere, and the Perelomov type coherent states of SO$(D+1)$ can be conveniently expressed in the harmonic function formulation. Based on this formulation, we studied the general properties of Perelomov type coherent states of SO$(D+1)$ in section 3, e.g. the peakedness property and the inner product. These properties made sure that we can define geometric operators using their classical expressions as $\mathbf{P}$ symbols. We also considered the properties of the matrix element functions on SO$(D+1)$ which are selected by Perelomov type coherent states, and we showed the peakedness property of these functions and proved that they can be regarded as the delta function on $Q_{D-1}$ in the large $N$ limit. Besides, we discussed the properties of the derivative of these matrix element function and conclude it as a Theorem in the final part of section 3.

Through studying the matrix element functions on SO$(D+1)$ selected by the Perelomov type coherent state, the properties of the spin-network states whose vertices are labelled by simple coherent intertwiners becomes clear.  For this kind of spin-network state, by showing that the expectation values of flux operators acting on vertices only involve the coherent intertwiner, we applied the peakedness properties of Perelomov coherent state of SO$(D+1)$ to prove that the simple coherent intertwiner weakly solves the quantum vertices simplicity constraint in large $N$ limit. Also, the properties of the matrix element functions on SO$(D+1)$ allowed us to calculate the expectation value of the standard D-volume operator (constructed directly from fluxes) in all dimensional LQG with $D$ being odd. By using the Theorem in section 3 and the Lemma in section 4, we argued that the expectation value of the D-volume operator with respect to the gauge invariant simple coherent intertwiner states can be given by replacing the operator $\hat{Q}$ in the expression of volume operator by the expectation value of $\hat{Q}$ with respect to the corresponding gauge fixed simple coherent intertwiner state, with some error terms which tends to zero in the large $N$ limit. In fact, the procedures for calculating the expectation value of the volume operator can be extended to other spatial geometric operators which are composed of flux operators when the Theorem in section 3 and the Lemma in section 4 are applicable.

In fact, coherent states are widely used in standard (1+3)-dimensional LQG and in particular allow to study the theory in a certain large quantum number limit where it behaves approximately classical at kinematical level, see e.g. \cite{ThiemannComplexifierCoherentStates}. Due to the formulation of standard (1+3)-dimensional LQG as an SU$(2)$ gauge theory, investigations using coherent states were mostly restricted to Perelomov type with group SU$(2)$ or Hall-Thiemann type \cite{ThiemannComplexifierCoherentStates}. In this paper, we investigated the Perelomov coherent state of SO$(D+1)$ which serves as the semiclassical state for spatial geometry based on its property that it minimalizes the quantum uncertainty of flux operators. However, there is still the quantum non-commutative bracket between holonomy operator and flux operator, whose quantum uncertainty can not be minimalized by the Perelomov coherent state of SO$(D+1)$. The one which is expected to achieve this goal is the Hall-Thiemann type coherent state of SO$(D+1)$. With the encourage of the previous works on SU$(2)$ coherent states \cite{Bianchi:2009ky}\cite{Calcinari:2020bft}, we hope that the properties of Perelomov coherent state of SO$(D+1)$ investigated in this paper will be helpful to the further study of the Hall-Thiemann type one.

\section*{Acknowledgments}
This work is supported by the National Natural Science Foundation of China (NSFC) with Grants No. 11775082, No. 11875006 and No. 11961131013. NB was supported by an International Junior Research Group grant of the Elite Network of Bavaria.

\appendix
\section{An error estimation}
Let us recall the Eqs. \eqref{predelta}, \eqref{deltaVV}. We notice that the error of the result \eqref{deltaVV} is given by two parts, which come from the two terms on the right hand side of Eq. \eqref{predelta} respectively. They are given by
\begin{equation}
\mathcal{E}_1\sim\epsilon ,\qquad \mathcal{E}_2\sim \left.\dim(\mathfrak{H}^N_{D+1})\left(\frac{(\cos\Delta\theta_1+\cos\Delta\theta_2)^2}{4}\right)^N\right| _{\Delta\theta_1\sim\Delta\theta_2\sim\epsilon\rightarrow0},
\end{equation}
where $\epsilon$ is the ``width'' of the region $\Delta$.
Denote $\frac{1}{e^\alpha}\equiv\cos^{2N}\epsilon$, we have in the limit $\epsilon\rightarrow0$
\begin{equation}
\frac{1}{e^\alpha}=(1-\sin^2\epsilon)^N\approx(1-\epsilon^2)^N,
\end{equation}
and
\begin{eqnarray}
-\alpha&\approx&N\ln(1-\epsilon^2)\\\nonumber
   &\approx&-N\epsilon^2.
\end{eqnarray}
Suppose $\epsilon=N^{-\frac{\beta}{2}}$ with $\beta>0$. Then, we have $\alpha\approx N^{(1-\beta)}$. Notice that $\dim(\mathfrak{H}^N_{D+1})=\frac{(N+D-2)!(2N+D-1)}{(D-1)!N!} \stackrel{ N\text{ large}}{\sim} N^{(D-1)}$ and suppose $\mathcal{E}_2=N^{-\rho}$, then we have
\begin{equation}
\mathcal{E}_2=N^{-\rho}\sim\frac{N^{(D-1)}}{e^{N^{(1-\beta)}}}.
\end{equation}
Taking natural logarithms on both sides, we get
\begin{equation}
\rho\ln N\sim N^{(1-\beta)}-(D-1)\ln N,\Rightarrow \rho\sim\frac{N^{(1-\beta)}}{\ln N}-(D-1)\sim(1-\beta)N^{(1-\beta)}-(D-1)
\end{equation}
in the limit $N\rightarrow\infty$. Now the total error can be estimated by
\begin{equation}
\mathcal{E}=\mathcal{E}_1+\mathcal{E}_2\sim N^{-\frac{\beta}{2}}+N^{((D-1)-(1-\beta)N^{(1-\beta)})}.
\end{equation}
It is easy to see that for a proper choice of $1>\beta>0$, i.e. $\beta=\frac{1}{2}$, the error will be $\mathcal{E}=N^{-\frac{1}{4}}+N^{((D-1)-\frac{1}{2}N^{\frac{1}{2}})}$, which tends to zero in large $N$ limit.
\section{The calculation of Eq. \eqref{epQ}}
Based on Eqs. \eqref{NgNgN} and \eqref{Qv}, we can express $\langle\hat{Q}_{v,\gamma}^2\rangle$ as
\begin{eqnarray}\label{epQ2}
&&\langle\hat{Q}_{v,\gamma}^2\rangle\\\nonumber
&=& \frac{\langle\gamma, (\vec{N},\vec{V})_v,...|\hat{Q}_{v,\gamma}^2|\gamma, (\vec{N},\vec{V})_v,...\rangle} {\langle\gamma, (\vec{N},\vec{V})_v,...|\gamma, (\vec{N},\vec{V})_v,...\rangle} \\\nonumber
&=&\frac{\sum_{\{e\}}\sum_{\{e\}'}s(\{e\})s(\{e\}')\int dg\epsilon_{...I J...KL...}\epsilon_{...I' J'...K'L'...}F_{\text{vol}}^{...IJ...KL...I'J'...K'L'...}(g)} {4(D!)^24^D\langle \vec{N},\vec{V}||\vec{N},\vec{V}\rangle}
\end{eqnarray}
with
\begin{eqnarray}
 &&F_{\text{vol}}^{...IJ...KL...I'J'...K'L'...}(g)\\\nonumber
 &:=& \prod_{\imath\in\{\imath|e_\imath\in\{e\}\cap\{e\}' \}}\langle N_\imath,V_\imath|X^{I J}X^{I' J'}g|N_\imath,V_\imath\rangle\\\nonumber
  && \times \prod_{\imath\in\{\imath|e_\imath\in\{e\}\ \text{and}\ e\notin\{e\}' \}}\langle N_\imath,V_\imath|X^{KL}g|N_\imath,V_\imath\rangle\\\nonumber
  && \times \prod_{\imath\in\{\imath|e_\imath\in\{e\}'\ \text{and}\ e\notin\{e\} \}}\langle N_\imath,V_\imath|X^{K'L'}g|N_\imath,V_\imath\rangle\\\nonumber
  && \times \prod_{\imath\in\{\imath|e_\imath\notin\{e\}'\ \text{and}\ e\notin\{e\} \}}\langle N_\imath,V_\imath|g|N_\imath,V_\imath\rangle.
\end{eqnarray}
Now the Theorem introduced in the final part of section 3 is applicable for $F_{\text{vol}}^{...IJ...KL...I'J'...K'L'...}(g)$. With this property, we can immediately get
\begin{eqnarray}
 &&\frac{F_{\text{vol}}^{...IJ...KL...I'J'...K'L'...}(g)}{ (\prod_{e_\imath\in\{e\}} N_{e_\imath})(\prod_{e_\jmath\in\{e\}'}N_{e_\jmath})}\\\nonumber
 &=&\prod_{\imath\in\{\imath|e_\imath\in\{e\}\cap\{e\}' \}}\left(V_\imath^{IJ}V_\imath^{I'J'}\langle N_\imath,V_\imath|g|N_\imath,V_\imath\rangle+\mathcal{O}^{IJI'J'}_{N_\imath}(g)\right)\\\nonumber
  && \times \prod_{\imath\in\{\imath|e_\imath\in\{e\}\ \text{and}\ e\notin\{e\}' \}}\left(V_\imath^{KL}\langle N_\imath,V_\imath|g|N_\imath,V_\imath\rangle+\mathcal{O}^{KL}_{N_\imath}(g)\right)\\\nonumber
  && \times \prod_{\imath\in\{\imath|e_\imath\in\{e\}'\ \text{and}\ e\notin\{e\} \}}\left(V_\imath^{K'L'}\langle N_\imath,V_\imath|g|N_\imath,V_\imath\rangle+\mathcal{O}^{K'L'}_{N_\imath}(g)\right)\\\nonumber
  && \times \prod_{\imath\in\{\imath|e_\imath\notin\{e\}'\ \text{and}\ e\notin\{e\} \}}\langle N_\imath,V_\imath|g|N_\imath,V_\imath\rangle
\end{eqnarray}
where $\mathcal{O}^{IJI'J'}_{N_\imath}(g)$,  $\mathcal{O}^{KL}_{N_\imath}(g)$ and $\mathcal{O}^{K'L'}_{N_\imath}(g)$ tend to zero in large $N_\imath$ limit. Then, the integral in Eq.\eqref{epQ2} can be given by
\begin{eqnarray}
&&\int dg\epsilon_{...I J...KL...}\epsilon_{...I' J'...K'L'...}F_{\text{vol}}^{...IJ...KL...I'J'...K'L'...}(g)\\\nonumber
&\stackrel{\vec N\text{ large}}{=}&4^D (\prod_{e_\imath\in\{e\}} N_{e_\imath})(\prod_{e_\jmath\in\{e\}'}N_{e_\jmath})\epsilon(\{e\})\epsilon(\{e\}')\langle \vec{N},\vec{V}||\vec{N},\vec{V}\rangle
\end{eqnarray}
with $\epsilon(\{e\}):=\epsilon_{I J...KL}V_{e_1}^{IJ}...V_{e_D}^{KL}|_{e_1,...,e_D\in\{e\}}$ and $\epsilon(\{e\}'):=\epsilon_{I J...KL}V_{e_1}^{I'J'}...V_{e_D}^{K'L'}|_{e_1,...,e_D\in\{e\}'}$. Finally, we get the result mentioned in section 4, which is
\begin{eqnarray}
\langle\hat{Q}_{v,\gamma}^2\rangle
&\stackrel{\vec N\text{ large}}{=}&\frac{\sum_{\{e\}}\sum_{\{e\}'}s(\{e\})s(\{e\}') (\prod_{e_\imath\in\{e\}} N_{e_\imath})(\prod_{e_\jmath\in\{e\}'}N_{e_\jmath})\epsilon(\{e\})\epsilon(\{e\}')} {4(D!)^2}\\\nonumber
&\stackrel{\vec N\text{ large}}{=}& \left(\frac{\langle\gamma, [\vec{N},\vec{V}]_v,...|\hat{Q}_{v,\gamma}|\gamma, [\vec{N},\vec{V}]_v,...\rangle} {\langle\gamma, [\vec{N},\vec{V}]_v,...|\gamma, [\vec{N},\vec{V}]_v,...\rangle}\right)^2=:([\hat{Q}_{v,\gamma}])^2.
\end{eqnarray}
Similar calculation can be done to give Eq.\eqref{epQ4}. Notice that we are interested in the expectation value of $\mathbb{P}_v^{\text{s}}\hat{Q}_{v,\gamma}^2\mathbb{P}_v^{\text{s}}$ originally. Thus we may consider the difference between the expectation value $\langle\hat{Q}_{v,\gamma}^4\rangle$ and $\langle\hat{Q}_{v,\gamma}^2\mathbb{P}_v^{\text{s}}\hat{Q}_{v,\gamma}^2\rangle$, which requires us to consider the non-diagonal element of $\hat{Q}_{v,\gamma}^2$ in coherent intertwiner space, that is
\begin{eqnarray}\label{non-diagonal}
 &&\frac{|\langle\gamma, (\vec{N},\vec{V}')_v,...|\hat{Q}_{v,\gamma}^2|\gamma, (\vec{N},\vec{V})_v,...\rangle|^2} {\langle\gamma, (\vec{N},\vec{V})_v,...|\gamma, (\vec{N},\vec{V})_v,...\rangle\langle\gamma, (\vec{N},\vec{V}')_v,...|\gamma, (\vec{N},\vec{V}')_v,...\rangle}\\\nonumber
 &\stackrel{\vec N\text{ large}}{=}&([\hat{Q}_{v,\gamma}])^4\frac{|\langle\vec{N},\vec{V}'||\vec{N},\vec{V}\rangle|^2} {\langle\vec{N},\vec{V}||\vec{N},\vec{V}\rangle\langle\vec{N},\vec{V}'||\vec{N},\vec{V}'\rangle},
\end{eqnarray}
where $|\gamma, (\vec{N},\vec{V})_v,...\rangle$ and $|\gamma, (\vec{N},\vec{V}')_v,...\rangle$ are only distinguished by the difference between the simple coherent intertwiner $||\vec{N},\vec{V}\rangle_v$ and the coherent intertwiner $||\vec{N},\vec{V}'\rangle_v$, and we follow the same procedures for calculating $\langle\hat{Q}_{v,\gamma}^2\rangle$ and use the Theorem in section 3 additionally. It is easy to see Eq.\eqref{non-diagonal} is a relatively small quantity compared with the diagonal element $\langle\hat{Q}_{v,\gamma}^2\rangle^2$ in large $\vec{N}$ limit if $\vec{V}$ and $\vec{V}'$ are not identical up to a global $SO(D+1)$ rotation. Hence we have $\mathbb{P}_v^{\text{s}}\hat{Q}_{v,\gamma}^2|\gamma, (\vec{N},\vec{V})_v,...\rangle\stackrel{\vec N\text{ large}}{=}\hat{Q}_{v,\gamma}^2|\gamma, (\vec{N},\vec{V})_v,...\rangle$ and then
\begin{equation}
\langle\hat{Q}_{v,\gamma}^4\rangle\stackrel{\vec N\text{ large}}{=}\langle\hat{Q}_{v,\gamma}^2\mathbb{P}_v^{\text{s}}\hat{Q}_{v,\gamma}^2\rangle.
\end{equation}
\bibliographystyle{unsrt}

\bibliography{Perelomovtypecoherentstates}

\end{document}